\documentclass[superscriptaddress,twocolumn,showkeys, amssymb,amsmath,nobibnotes,aps,prd, nofootinbib]{revtex4-2}
\pdfoutput=1
\usepackage{graphicx,subfigure,bm,color,psfrag,hyperref}
\usepackage{amsfonts}
\usepackage{lipsum}
\usepackage{mathtools}
\usepackage{verbatim}
\usepackage{colortbl}
\usepackage[table]{xcolor}

\hypersetup{colorlinks,linkcolor={blue},citecolor={cyan},urlcolor={cyan}}

\begin{document}

\title{Is Dark Energy an Effective Manifestation of Non-equilibrium Thermodynamics? -- Insights from DESI}

\author{Sauvik Bhattacharjee}
\email{sauvikb\textunderscore r@isical.ac.in}
\affiliation{Physics and Applied Mathematics Unit, Indian Statistical Institute, 203 B. T. Road, Kolkata 700108, India}

\author{Sudip Halder}
\email{sudip.rs@presiuniv.ac.in}
\affiliation{Department of Mathematics, Presidency University, 86/1 College Street, Kolkata 700073, India}

\author{Jaume de Haro}
\email{jaime.haro@upc.edu}
\affiliation{Departament de Matem\`atiques, Universitat Polit\`ecnica de Catalunya, Diagonal 647, 08028 Barcelona, Spain}

\author{Supriya Pan}
\email{supriya.maths@presiuniv.ac.in}
\affiliation{Department of Mathematics, Presidency University, 86/1 College Street, Kolkata 700073, India}
\affiliation{Institute of Systems Science, 
Durban University of Technology, Durban 4000, Republic of South Africa}

\author{Emmanuel N. Saridakis }\email{msaridak@noa.gr}
\affiliation {National Observatory of Athens, Lofos Nymfon, 11852 Athens,
Greece}
\affiliation {Departamento de Matem\'{a}ticas, Universidad Cat\'{o}lica del
Norte, Avda. Angamos 0610, Casilla 1280 Antofagasta, Chile}
\affiliation { CAS Key Laboratory for Researches in Galaxies and Cosmology,
Department of Astronomy, University of Science and Technology of China, Hefei,
Anhui 230026, P.R. China}

\begin{abstract}

We investigate the background cosmological expansion on the onset of cosmological homogeneous matter creation scenario, a dynamical dark matter approach ($w_{\rm dm} \neq 0$), and an alternative 
approach to both dark energy and modified gravity theories,  after the recent 
DESI DR2-BAO release. We consider that the total matter sector consists of 
three independently evolving components, namely, radiation, baryons, and dark 
matter, with the latter being governed by an adiabatic matter creation process, affects the background homogeneously,  
 leads to a modified continuity equation.  Though the total stress-energy tensor is conserved the only violation of the conservation law in the dark matter sector is coming from the creation pressure, and under a proper choice of dark-matter particle   
creation rate one can obtain 
the present accelerating phase as well as the past thermal history of the 
Universe. We study two specific matter creation rates. 
By applying the  dynamical-system analysis we show that both
   Model I and Model II  can mimic a   
$\Lambda$CDM-like behavior.
 Furthermore,    we perform  a detailed observational confrontation 
using   a series of latest observational datasets including Cosmic Chronometers 
(CC), Supernovae Type Ia (SNIa) 
(Pantheon+, DESY5 and Union3 samples) and DESI Baryon Acoustic Oscillations 
(BAO) (DR1 and DR2 samples). In both Model I and Model II
we find   evidence of matter creation   
  at many standard deviations. 
 Finally, applying the AIC and BIC information criteria we find that  
 Model I is statistically equivalent with  $\Lambda$CDM scenario, while  Model 
II  shows a mixed picture, namely for most datasets  $\Lambda$CDM scenario 
is favoured, however when  DESI data are included    matter 
creation Model II is favoured  over   $\Lambda$CDM paradigm.

\end{abstract}

\maketitle

\section{Introduction}
\label{sec-introduction}

Although the concordance $\Lambda$-Cold Dark Matter ($\Lambda$CDM) paradigm 
is observationally supported at a high level,   it is  facing theoretical and 
observational  challenges \cite{DiValentino:2021izs,Perivolaropoulos:2021jda,Schoneberg:2021qvd,Abdalla:2022yfr,Kamionkowski:2022pkx,CosmoVerse:2025txj}. 
As alternatives,  one can  either introduce extra fluids/particles 
with negative pressure within the context of general relativity (GR), resulting 
to the concept of  dark energy (DE) ~\cite{Peebles:2002gy,Copeland:2006wr,Bamba:2012cp} 
(see also \cite{Perlmutter:1999jt,Boisseau:2000pr,Amendola:1999er,Parker:1999td, Caldwell:1999ew,Kamenshchik:2001cp, Padmanabhan:2002cp, Cardenas:2002np, Khoury:2003rn, Khoury:2003aq, Li:2004rb, Wang:2004cp, Cai:2009zp, Uggla:2013uqa, Paul:2013sha, Sharov:2015ifa, Leanizbarrutia:2017afj, Lombriser:2018olq, Pan:2020mst,  Landim:2021ial, Artymowski:2021fkw, Yang:2021hxg, Oriti:2021rvm, Teixeira:2022sjr,  Lee:2022cyh, Ballardini:2023mzm, daCosta:2023mow, Forconi:2023hsj, vanderWesthuizen:2023hcl, Cline:2023cwm, Patil:2023rqy, Giare:2024gpk, Notari:2024rti, Giare:2024smz,  Christiansen:2024hcc, Silva:2025hxw, Sabogal:2025mkp}), 
or one can modify GR in various ways, obtaining
modified gravity (MG) \cite{Nojiri:2006ri,DeFelice:2010aj,Clifton:2011jh, Cai:2015emx, 
Nojiri:2017ncd, Bahamonde:2021gfp,Heisenberg:2023lru}.  
 (also see \cite{Nojiri:2003ft,Dolgov:2003px,Carroll:2004de,Cognola:2006eg,Ferraro:2006jd,Apostolopoulos:2007cr,Sotiriou:2008it,Harko:2008qz,Bamba:2008hq,Elizalde:2010ep,Tsujikawa:2010zza,Martinelli:2010wn,Harko:2011kv,Martinelli:2011wi,Borowiec:2011wd,Li:2011vk,Brax:2011aw,Mak:2011bt,Basilakos:2012uu,DiValentino:2012yg,Gao:2012fd,Briscese:2012ys,Clemson:2012im,Chakraborty:2012kj,Brax:2013fda,Leonard:2015hha,Nunes:2016qyp,Nunes:2016drj,Nunes:2018evm,Samart:2021viu, Lymperis:2022oyo, Santos:2022atq,Kumar:2023bqj,Ferrari:2025egk,Ayuso:2025vkc,Yang:2024tkw,Xia:2024tps}). 

However, one interesting third avenue and  possible alternative to both DE and  MG, is  to construct cosmological scenarios equipped with gravitationally induced adiabatic matter creation 
\cite{Steigman:2008bc,Lima:2008qy,Lima:2009ic,Lima:2011hq,Jesus:2011ek,Lima:2012cm,Lima:2014qpa,Lima:2014hda,Chakraborty:2014fia,Lima:2015xpa,Baranov:2015eha,deHaro:2015hdp,Pan:2016jli,Paliathanasis:2016dhu,Pan:2018ibu,Elizalde:2024rvg,Kuzmin:1998uv,Ema:2018ucl,Shiu:2003ta,Ema:2019yrd,Modak:2019jbg}. In particular, it is possible to 
generate a  negative ``creation pressure'' through the produced particles from 
a matter creating field, and as a result   this creation pressure could 
drive the observed accelerated expansion of the universe. Also one can categorize the outcome of this model into a class of dynamical dark matter~\cite{Muller:2004yb,Armendariz-Picon:2013jej, Gariazzo:2017pzb, Kopp:2018zxp} where dark matter shows non-cold signatures.

In this work we are interested in constructing a scenario with 
 the production of a dark 
fluid  corresponding to   cold dark matter, with a phenomenological rate. A 
bare choice of production rate could be dependent on the existing density 
parameter of the dark sector and the Hubble parameter. In our analysis we are 
adopting a choice of adiabatic gravitational matter creation theory from the 
earlier works of  Lima et.al. \cite{Lima:2009ic,Lima:2011hq,Lima:2012cm} 
in which   the idea of ``Creation Cold Dark Matter'' or ``CCDM'' 
cosmology was introduced. The 
underlying thermodynamical process creates an instantaneous negative pressure, 
which   primarily describes the universe acceleration at present epoch.
Concerning the total energy density this model 
behaves like the background $\Lambda$CDM cosmology. Concerning the 
matter-creation rate,  the form  $\Gamma  = 3 \alpha H 
\left(\frac{\rho_{\rm c,0}}{\rho_{\rm dm}}\right)$, known as the LJO model (introduced by Lima, Jesus, and Oliveira),  was used in several works 
\cite{Lima:2009ic,Lima:2011hq,Lima:2012cm}. However, in our analysis  we will consider  the baryonic and radiative 
contributions into the Friedmann equations too.

The plan of the work is as follows: In Section \ref{sec-matter-creation} we   
present the thermodynamic conditions of matter creation in the 
maximally symmetric and expanding space-time,   
introducing the concept of instantaneous creation pressure which 
causes the accelerating expansion.  In 
Section \ref{sec-dynamical-models}  we formulate a dynamical system with 
the density parameters and  we investigate the dynamics in the phase space 
for Model I and Model II. In \ref{sec-data} we  
constrain the model parameters using  Cosmic Chronometers (CC), Type Ia Supernovae (SNIa) and DESI-BAO  
latest astronomical datasets. In \ref{sec-results} we enlist the results of data analysis. Finally, \ref{sec-summary} comprises the 
insights and a comparative discussions for both the dynamical analysis and data 
analysis of the matter creation models.

\section{Matter Creation Cosmologies: a brief overview}
\label{sec-matter-creation}

  We consider a flat Friedmann-Lema\^{i}tre-Robertson-Walker (FLRW) line 
element given by
\begin{align}
ds^{2}=-dt^{2}+a^{2}(t)\delta_{ij}dx^{i}dx^{j},
\label{eq1} 
\end{align}
where $a(t)$ is the expansion scale factor.
Additionally, we  consider that the gravitational sector is 
described by general relativity and that   matter   is minimally 
coupled to gravity. Concerning  matter, we assume that it is comprised of 
relativistic radiation, non-relativistic baryons, and a pressure-less dark 
matter.  

We consider  a 
weakly interacting ideal fluid whose  stress-energy tensor is given by 
\begin{equation}
    T^{\mu \nu}=(\rho + p)u^{\mu}u^{\nu} + p g^{\mu \nu}\label{eq2},
\end{equation}
where $\rho$ and $p$ are respectively the energy density and pressure of the 
fluid  under consideration, $u^{\mu}$ is its four velocity, and 
$g^{\mu\nu}$ is the metric tensor of the desired spacetime. Therefore, one can 
  identify that $T^{00}=\rho$ and $T^{1}_{1}=T^{2}_{2}=T^{3}_{3}=p$. 
Furthermore,  the conservation of stress-energy tensor gives    $\nabla_{\nu} 
T^{\mu \nu}=0$.
In the context of matter creation, since we have  deviation from   
thermodynamic equilibrium, we can split the pressure in a general way as  
\begin{equation}
    p=p_{eq}+\delta p\label{eq3},
\end{equation}
where $p_{eq}$ is the equilibrium pressure  and $\delta p$ is the correction 
term for the sudden in-equilibrium phase. Eq.~\eqref{eq3} and the 
conservation equation of stress-energy tensor yields a modified version of 
energy density continuity equation, namely 
\begin{equation}
  \partial_{t}\rho + 3H(\rho+p_{eq}+\delta 
p)=0.
\end{equation}

   As we   mentioned above, in the matter-creation content  a particular 
kind of matter\footnote{We are agnostic about the underlying microphysical origin of particle production and instead focus on its macroscopic consequences.}
(which is weakly interacting and described as ideal fluid) 
is being created from the gravitational system.  Its particle flux vector and 
entropy 
flux vector are defined respectively as
\begin{eqnarray}
    N^{\mu}=& n u^{\mu},\label{eq4} \\
    S^{\mu}=& \sigma n u^{\mu}\label{eq5},
\end{eqnarray}
where $n$ is the co-moving particle number density and $\sigma$ is the entropy 
per particle in a co-moving volume.
The creation of particles immediately yields
\begin{equation}
    \nabla_{\mu} N^{\mu}= \partial_{t}n+3Hn \geq 0\label{eq06},
\end{equation}
whereas from the second law of thermodynamics we have 
\begin{equation}
    \nabla_{\mu} S^{\mu}=\nabla_{\mu} N^{\mu}\sigma +  n\dot{\sigma}\geq 
0\label{eq7},
\end{equation}
where $H \equiv \dot{a}/a$ is the Hubble rate, and overdots denote cosmic-time 
derivatives.
We assume a volume $V$ which contains $N$ 
 weakly interacting particles of the same kind. Then, the first law of  
thermodynamics is written as
\begin{equation}   d\mathcal{E} = dQ - pdV \label{eq8}.
 \end{equation}
 Now if we allow for   particle number change,  by anticipating the 
creation of particles, the above equation becomes
\begin{equation}
     d\mathcal{E} = dQ - pdV + \frac{\rho +p}{N}VdN, \label{eq9}
\end{equation}
  known also as the Gibbs-Duhem relation. Since in our analysis we have 
considered adiabatic particle creation, the above relation finally becomes
\begin{equation}
    d(\rho V)+pdV- \frac{\rho +p}{n}dN=0 \label{eq10},
\end{equation}
where $\mathcal{E}=\rho V$ is the system energy  and thus the aforementioned  
particle number density is expressed as $n=\frac{N}{V}$. Since   the third term 
in     \eqref{eq9} represents the heat received by the system due to the 
change in particle number, we deduce that matter creation    acts as a 
source of internal energy. 
 Finally, since   $\delta p$ in (\ref{eq3}) arises from the creation pressure, 
and  since we are interested in adiabatic matter creation,
from now on we rename $\delta p$ as $p_{\rm c}$, termed as the creation pressure.

Incorporating  relations  (\ref{eq06}) and (\ref{eq8})   we  acquire
\begin{equation}
   \nabla_{\mu} S^{\mu}= - \frac{p_{\rm c}\nabla_{\mu}u^{\mu}}{T}-\left( \frac{\rho 
+p}{nT}-\sigma\right)\nabla_{\mu}N^{\mu} \label{eq11}.
\end{equation}
 Though the processes are  isentropic in global sense the particle creation 
leads to heat and entropy generation into the system. In Ref.~\cite{Calvao:1991wg} 
the 
authors proposed a phenomenological ansatz for the creation pressure, namely
 \begin{equation}
     p_{\rm c}= -\alpha \frac{\nabla_{\mu}N^{\mu}}{\nabla_{\mu}u^{\mu}}\label{eq12},
 \end{equation}
 with $\alpha$ a parameter,
 which implies that if there is no particle creation (i.e. 
$\nabla_{\mu}N^{\mu}=0$), then the  creation pressure   and entropy production 
  $\nabla_{\mu}S^{\mu}$ are both     zero.
 From Eq.~\eqref{eq06}, Eq.~\eqref{eq11}  and Eq.~\eqref{eq12} we obtain 
 \begin{equation}
     \dot{\sigma} = \frac{\nabla_{\mu}N^{\mu}}{nT}\left( \alpha - 
\frac{\rho+p}{n}\right)\label{eq13} .
 \end{equation}
 Since we   restrict in   adiabatic processes, 
 \eqref{eq13}  implies that $\alpha=1$, and thus 
 we can rewrite \eqref{eq12} as
\begin{eqnarray}
  p_{\rm c}=\left(\frac{\rho+p}{n}\right) \frac{\partial_{t} n 
+3Hn}{3H}\label{eq14} 
     \end{eqnarray}
     or
\begin{eqnarray}  \partial_{t} n +3Hn = \frac{3Hn}{\rho+p}p_{c}\label{eq15}.
\end{eqnarray}
We are considering the adiabatic particle creation for dark matter species only. Hence, one can consider a simpler choice of the creation pressure of dark 
matter, namely
\begin{eqnarray}\label{eqn:pc}
    p_{\rm c} = -\frac{\Gamma}{3H} \left(p_{\rm dm} + \rho_{\rm dm} \right) = 
-\frac{\Gamma}{3H} \rho_{\rm dm}, \label{crsn_press}
\end{eqnarray}
where dark matter is assumed to be pressure-less. Therefore,  
inserting \eqref{crsn_press} into \eqref{eq15} yields
 \begin{equation}
  \partial_t n + 3Hn = \Gamma n \label{eq17}   .
 \end{equation}
 
Since radiation and baryonic matter  are not involved in any creation 
processes, they  obey a barotropic equation of state, namely
$p_{i}=w_{i}\rho_{i}$, where  $\rho_{i}$ and $p_{i}$ are the pressure 
and energy density  of the corresponding fluid respectively, with $w_i$ its 
equation-of-state parameter (for baryons and radiation we have $w_b = 0$ and 
$w_r = 1/3$). We mention that we do not consider any interaction between 
the various sectors, and thus the continuity equations for radiation and 
baryonic matter
are the standard ones.  On the 
contrary, for dark matter, the co-moving density varies over time due to the 
adiabatic creation (apart from the usual Hubble expansion). In summary, 
 the continuity equations 
for the three fluids  are written as
\begin{eqnarray}
   && \dot{\rho}_{\rm r}  + 4H\rho_{\rm r} = 0 ,\label{eq18} \\
   && \dot{ \rho}_{\rm b} + 3H\rho_{\rm b} = 0,\label{eq19}  \\
   &&  \dot{\rho}_{\rm dm} + 3H\rho_{\rm dm} = \Gamma\rho_{\rm dm} \label{eq20}.
\end{eqnarray}
Eq.~\ref{eq20} can be rewritten as the following equation 
\begin{equation}
     \dot{\rho}_{\rm dm} + 3H\rho_{\rm dm}\left(1-\frac{\Gamma}{3H}\right) = 0, \label{ddm}
\end{equation}
which is equivalent to the continuity equation of DM with a dynamical equation of state, $w_{\rm dm}= -\frac{\Gamma}{3H}$.  
As usual,  in a spatially flat FLRW universe, the two Friedmann 
equations   are  
\begin{eqnarray}
&& H^2  = \frac{\kappa^2}{3}(\rho_{\rm dm}+\rho_{\rm b}+\rho_{\rm r}),\label{friedmann-1A} \\
&&2\dot{H}+3H^2 =-{\kappa}^2(p_{\rm dm}+p_{\rm b}+p_{\rm r}+p_{\rm c}),\label{friedmann-2A}
\end{eqnarray}
where 
$\kappa^2$ is the    gravitational constant. Finally, as usual it is convenient 
to introduce the deceleration parameter 
\begin{eqnarray}
\label{decpar}
q=-1 - \frac{\dot{H}}{H^2}.
\end{eqnarray} 
 
Since we are interested in obtaining accelerated solutions, which are   
driven by matter creation, we need to consider a negative creation pressure  
($p_c<0$) in \eqref{friedmann-2A}. This immediately 
implies that the creation rate  $\Gamma$ must be  positive\footnote{We also note that $\Gamma$ must have the dimension of inverse of time, like the Hubble parameter, in other words, $\Gamma/H$ is a dimensionless quantity.}. Moreover, 
the creation rate   can be 
constant, but it is  more general and natural if it depends on the 
Hubble 
function,   on the total or critical density of the 
universe, or on  the dark-matter density.  Lastly, we consider
our model to be   extra-galactic and free from galactic thermodynamical 
fluctuations.

\section{Specific models and dynamical evolution}
\label{sec-dynamical-models}

In this section we will consider specific matter-creation models, and we will 
investigate in detail their dynamical evolution by performing a phase-space 
analysis. 

 In order to transform the cosmological equations into their autonomous form we
need to introduce    proper auxiliary variables. In the case of equations  
  (\ref{eq18})-(\ref{friedmann-1A}) this can be done 
through
\begin{eqnarray}\label{variables}\
x \equiv  \frac{\kappa^2 \rho_{\rm dm}}{3H^2},\quad  y \equiv  \frac{\kappa^2 \rho_{\rm r}}{3H^2}, \quad  \xi \equiv \frac{H_0}{H_0+H},
\end{eqnarray}
where, $x$, $y$, $\xi$,  and $H_0$  are  the dark matter density 
parameter ($\Omega_{\rm dm}$), the radiation density parameter ($\Omega_{\rm r}$), the dimensionless Hubble-based variable and 
the Hubble parameter at present epoch, respectively. Thus, the baryon density 
parameter could be expressed by \eqref{friedmann-1A} as  
$\Omega_{\rm b}=(1-x-y)$. 

Let us construct the autonomous system for a general creation rate. 
To achieve this  we consider   derivatives of the dynamical variables 
with respect to the   time variable $N=\ln(a/a_0)$, where   $a_0$ 
is the scale factor at present epoch set to 1 for simplicity. 
After some   algebra, the autonomous system for a general creation rate 
$\Gamma$ becomes 
\begin{subequations}\label{ODEs-Gen-Gamma}
\begin{eqnarray}
   && x' = xy + \frac{\Gamma}{H_0}x(1-x)\left(\frac{\xi}{1-\xi}\right) \label{eq5} \\
   && y' = y\left[y-1 - \frac{\Gamma}{H_0}x\left(\frac{\xi}{1-\xi}\right)\right] \label{eq6} \\
   && \xi' = \frac{3}{2} \xi (1-\xi) \left[1+ \frac{y}{3} - 
\frac{\Gamma}{3H_0}x\left(\frac{\xi}{1-\xi}\right) \right] \label{eq7}  .
\end{eqnarray}
\end{subequations}
The physical domain $\mathbf{R}$ of the dynamical system is defined by the 
Friedmann constraint~\eqref{friedmann-1A} and the definition of the $\xi$ 
variable, namely
\begin{eqnarray}\label{physical-region}
\mathbf{R} =\Bigl\{ (x,y,\xi) \in \mathbb{R}^3: 0\leq x\leq 1, 0\leq y \leq 1, 
\nonumber\\ 0\leq \xi\leq 1, x+y\leq 1 \Bigr\},
\end{eqnarray}
corresponding to  a prism-like 3D region in the phase space. 

As we mentioned above, due to Eq. (\ref{eq20}), in matter creation cosmologies 
dark matter energy density does not follow its standard ($\Lambda$CDM-case) 
form. Thus, a general solution of  (\ref{eq20}) can be written as
\begin{equation}\label{rho_trial}
    \rho_{\rm dm} =  \rho_{\rm dm,0}a^{-3}f(a),
\end{equation}
with $\rho_{\rm dm,0}$ is the dark matter 
 energy density at present and where $f(a)$ is any suitable continuous function 
of $a$ that does not correspond to a singular $\Gamma$. From  
(\ref{rho_trial}) and (\ref{eq20}) we acquire
\begin{eqnarray}\label{Gen-Gamma}
    \Gamma =  \frac{H }{f}   \frac{df}{dN}.
\end{eqnarray}
  Using the form of creation rate  \eqref{Gen-Gamma}  we can rewrite 
the dynamical system  \eqref{ODEs-Gen-Gamma} as 
\begin{subequations}\label{eq29}
\begin{align}
    x' =& x\left[\frac{1}{f}\frac{df}{dN}(1-x) +y \right], \\
    y' =& y\left(y-1 - \frac{1}{f}\frac{df}{dN}x \right),\\
    \xi'=& \frac{3}{2}\xi(1-\xi)\left(1+\frac{y}{3}- 
\frac{x}{3}\frac{1}{f}\frac{df}{dN} \right) .
    \end{align}
\end{subequations}
We mention here that for general forms of $\Gamma$ we cannot obtain analytical 
forms for  $f(a)$ and hence $\rho_{\rm dm}$. In the following subsections we 
consider specific forms of $\Gamma$, namely we examine specific matter-creation 
models.

 \begin{table*}[t]    
\centering
\begin{tabular}{|c|c|c|c|c|c|c|c|c|}
 \hline
   Fixed Points & $(x,y,\xi) $ & Eigenvalues & Stability & $\Omega_{\rm dm}$ & $\Omega_{\rm r}$  & $\Omega_{\rm b}$  & Acceleration \\
   \hline\hline
   $A_0$ &$\left(x_c,0,0\right)$ & $\left(0,1,\frac{3}{2}\right)$&  Unstable  & $x_c$ & 0 & $1-x_c$  &  $q=\frac{1}{2}$, No \\
   
 $A_1$ & $(1,0,1)$ &   $\left(-3\alpha, \frac{3\alpha}{2},-3\alpha\right)$&  Saddle & 1 & 0 & 0 & $q\to -\infty$, Yes \\
  
  $A_2$ &$ \left(1,0, \frac{1}{\sqrt{\alpha}+1}\right)$ & $\left(-\frac{3 \alpha}{(1 + \sqrt{\alpha})^2},
 -\frac{3 \alpha}{(1 + \sqrt{\alpha})^2},
-\frac{3 \alpha \sqrt{\alpha}+ 2 \alpha}{(1 + \sqrt{\alpha})^2}\right)$ &   \textbf{Stable} & 1 & 0 & 0 & $q=-1$, Yes \\

$A_3$ & $(0,1,0)$ & $(1,1,2)$&  Unstable   & 0 & 1 & 0 &$q=1$ , No \\
\hline
\end{tabular}
\caption{Summary of the critical points, their nature (in terms  of   
stability) and the corresponding cosmological parameters for   Model 
\hyperref[M1]{I}  characterized by the matter creation rate  $\Gamma = 3\alpha 
H \left(\rho_{\rm c,0}/\rho_{\rm dm} \right)$. }
\label{Table-M1}
\end{table*}

\subsection{Model I:  $\Gamma  = 3 \alpha H \left(\frac{\rho_{\rm c,0}}{\rho_{\rm  
dm}}\right)$}\label{M1}

As a first  Creation Cold Dark Matter (CCDM) model  we consider the following
matter creation rate \cite{Lima:2009ic}:    
\begin{equation}\label{eq30}
\Gamma  = 3 \alpha H \left(\frac{\rho_{\rm c,0}}{\rho_{\rm dm}}\right),
\end{equation}
where $\rho_{\rm c,0}$ is the critical  energy density at present and $\alpha$ is 
a positive parameter. The  inversely proportional dependence 
with the dark matter density implies a gradual reduction in the creation rate.
Choice  (\ref{eq30}) allows for an analytical solution of the matter 
density $\rho_{\rm dm}$  in (\ref{eq20}). In particular, using 
Eqs. (\ref{eq20}) and (\ref{eq30}) we extract  the solution as\footnote{The analytical solution is presented in Appendix~\ref{appen-A}} 
\begin{equation}
    \rho_{\rm dm} = ( \rho_{\rm dm,0}-\alpha  \rho_{\rm c,0})a^{-3} + \alpha 
\rho_{\rm c,0}. 
\label{m1-rho-dm-sol}
\end{equation}
Note that since the above solution naturally generates a constant term which resembles a cosmological constant term, allowing the model to reproduce the $\Lambda$CDM-like late-time acceleration without any need of a cosmological constant which is inserted in GR.  

 The  first Friedmann equation (\ref{friedmann-1A}) can be 
written as
{\small{
 \begin{equation}
    \left(\frac{H}{H_0}   \right)^2  = \alpha + (\Omega_{\rm dm,0}- \alpha)(1 + z)^3 + \Omega_{\rm b,0} (1+z)^3 + \Omega_{\rm r,0}(1+z)^4,
\label{m1_data}
\end{equation}}}
where the density parameter  for the $i$-th sector at present is given by 
  $\Omega_{\rm i,0}\equiv \frac{\rho_{\rm 
i,0}}{\rho_{\rm c,0}}$, and thus we have $\Omega_{\rm dm,0}+\Omega_{\rm 
b,0}+\Omega_{\rm r,0} =1$, while we have introduced
  the redshift  $ z$ through $1+z 
= 1/a$. Note that the above expansion law mimics the $\Lambda$CDM cosmological 
scenario including baryons and radiation.   

In this case, the autonomous system (\ref{ODEs-Gen-Gamma})   becomes 
\begin{subequations}\label{DS-1}
   \begin{align}
    x'=& xy(1-\xi)^2+3\alpha(1-x)\xi^2, \\  
    y'=& y\left[(y-1)(1-\xi)^2-3\alpha\xi^2\right], \\ 
    \xi'=& \frac{3}{2}\xi(1-\xi)\left[\left(1+\frac{y}{3}\right)(1-\xi)^2-\alpha 
\xi^2 \right],
\end{align} 
\end{subequations}
where we have multiplied the above equations by $(1-\xi)^2$ to handle the  singularity at $\xi=1$.  
 Additionally, the deceleration parameter (\ref{decpar}) for this case reads as 
\begin{equation}
    q= \frac{1}{2}\left[(1+y) - 3\alpha  \left(\frac{\xi}{1-\xi}\right)^2  
\right]. \label{q-M1}
\end{equation}

The above three-dimensional autonomous system exhibits three invariant 
planes, namely $y=0$, $\xi=0$ and $\xi=1$. On the other hand, we can show that 
$x'+y'=(x+y-1)\left(y(1-\xi)^2+3\alpha \xi^2\right)$, which implies that $x+y=1$ 
is also an invariant plane. Lastly, at $x=0$ we have $x'=3\alpha \xi^2$, 
and this positive 
gradient of $x$ implies that no trajectories will be entering the $x<0$ region 
by 
passing through $x=0$ plane in forward time. Hence, the
phase space is positively invariant. 
 
The critical points of the above system,  their stability and the corresponding 
cosmological parameters have been enlisted in Table \ref{Table-M1} (we 
have used  linear stability theory, complemented with numerical elaboration, 
throughout each fixed point to investigate their stability nature).  
Note that apart from  the critical point $A_2$, all other critical points lie 
upon either $\xi=0$ or $\xi=1$ plane.
As one can see, $\xi=0$ plane has $H \to \infty$, $\rho_{\rm dm} \to \infty$ and $\Gamma \to 0$,  while $\xi=1$ plane has $H \to 0$, $ \rho_{\rm dm} \to 0$ \ and $\Gamma \to \infty$. The $\xi =0$ and $\xi=1$ planes inevitably correspond to 
two limiting phases of cosmic dynamics. Let us describe the type and stability of each critical point separately.
(\texttt{i}) $\bf{A_{0}}$: This is a curve   containing 
non-isolated critical points. At these points, two of the eigenvalues of the 
Jacobian matrix are   positive   and one is zero, indicating that these are 
normally hyperbolic critical points. From the linear stability theory and the 
center-manifold theorem \cite{Leon:2014yua,Leon:2012mt}, we conclude that these 
are unstable points. They correspond to dark matter and baryon dominated, 
decelerated $(q>0)$ solutions. (\texttt{ii}) $\bf{A_{1}}$: At this fixed point   the Jacobian matrix has two 
negative and one  positive eigenvalues, and thus it is a saddle point. It 
corresponds to complete dark-matter dominated,  super-accelerated 
$(q<-1)$ solution. (\texttt{iii}) $\bf{A_{2}}$: All the three eigenvalues of the Jacobian matrix are 
negative at this critical point which makes it a stable point.  This point is 
dominated by dark matter and corresponds to an accelerating, de-Sitter 
$(q=-1)$, phase. Hence, this solution will attract the Universe at late times.
    (\texttt{iv}) $\bf{A_{3}}$:
    At this critical point, all eigenvalues are positive, and thus it is an 
unstable point. It corresponds to  a radiation-dominated 
decelerating phase. 
\begin{figure}[t]
    \centering
    \includegraphics[width=0.49\textwidth]{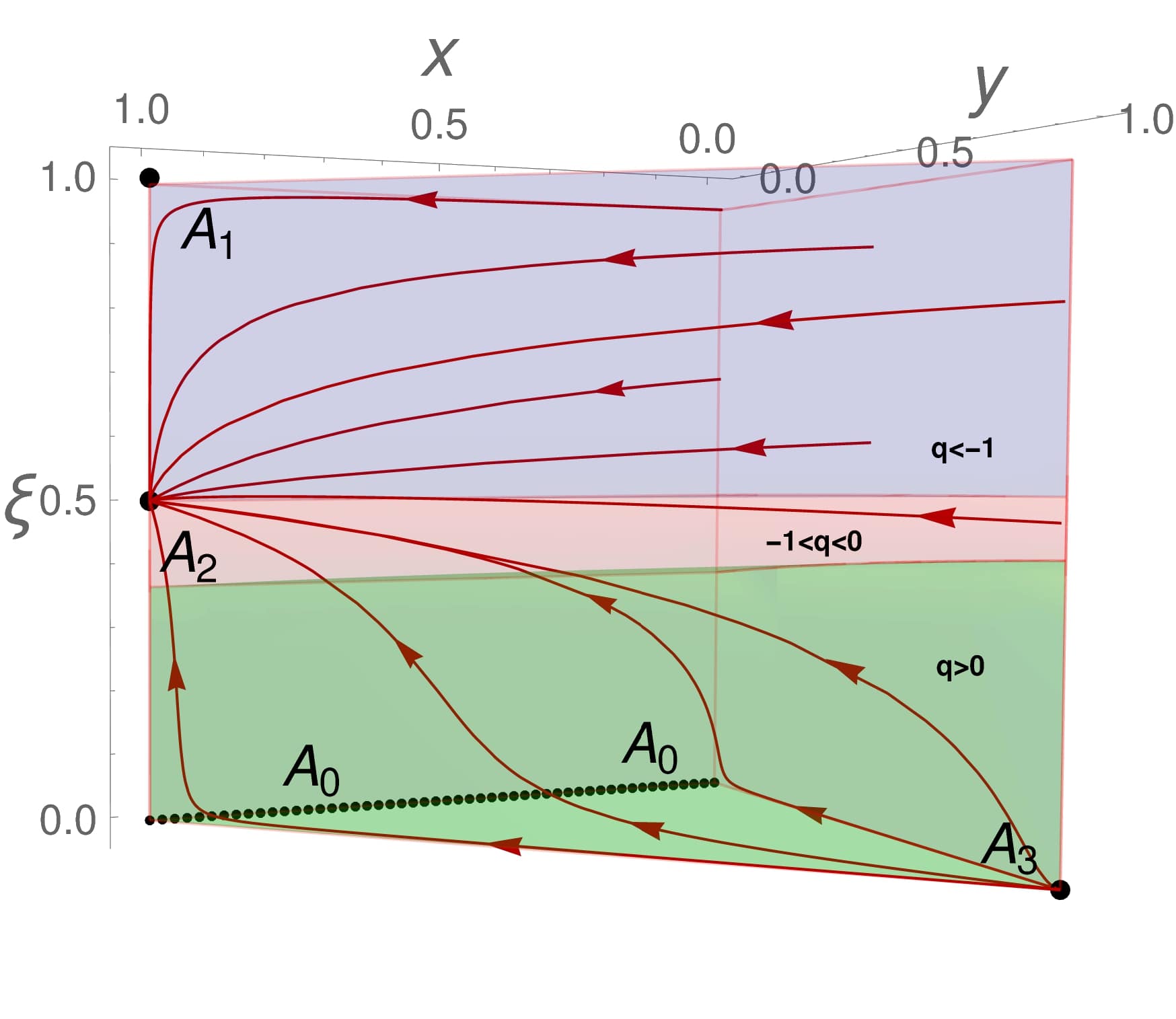}
    \caption{{\it{ The phase space diagram for Model \hyperref[M1]{I}, namely 
with 
  matter creation rate $\Gamma = 3\alpha H \left(\rho_{\rm c,0}/\rho_{\rm 
dm} \right)$, for  $\alpha=1$.    The green, pink and blue regions denote 
decelerated ($q>0$), accelerated ($-1<q<0$) and super-accelerated ($q<-1$) 
cosmic phases, respectively.  }}}
    \label{fig-(i)}
\end{figure}
\begin{figure}[t]
    \centering
    \includegraphics[width=0.9\linewidth]{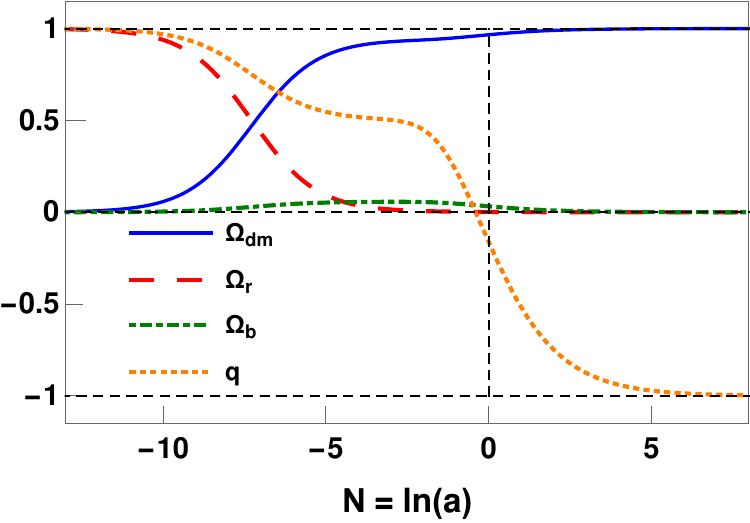}
    \caption{\textit{The evolution of the density parameters $\Omega_{\rm dm}$, 
$\Omega_{\rm r}$, $\Omega_{\rm b}$ and the deceleration parameter $q$ with 
respect to $\ln(a)$, for  Model \hyperref[M1]{I}, for 
$\alpha=1$, and imposing the initial conditions $x(-1)=0.896$,
$y(-1)=0.0007$, $\xi(-1)=0.278$.}}
    \label{fig-M1-Om-q-N}
\end{figure}

In  Fig.~\ref{fig-(i)}, we present some orbits in the phase space. As we 
observe,   one orbit starts from 
the close vicinity of $A_{3}$ (radiation dominated decelerated unstable point), 
then is repulsed by the closure with curve $A_{0}$ (DM+Baryon decelerated 
points) and finally is attracted by  $A_{2}$ (DM-dominated  accelerated 
solution). 
Additionally, in Fig.~\ref{fig-M1-Om-q-N} we depict  the evolution of the 
density 
parameters $\Omega_{\rm dm}$, $\Omega_{\rm r}$, $\Omega_{\rm b}$, as well as the 
deceleration 
parameter $q$ with respect to $\ln(a)$.
 Note that the initial conditions of $x$, $y$ and $\xi$ which are used for the evolution plot of Fig.~\ref{fig-M1-Om-q-N} are consistent  with the $\Lambda$CDM embedded Planck 2018 results~\cite{Planck:2018vyg}.\footnote{For the other evolution plots such as Fig.~\ref{fig-M2(i)a-Om-q-N}, Fig.~\ref{fig-M2(ii)-Om-q-N} and Fig.~\ref{fig-M2(iv)-Om-q-N}  the initial conditions have been taken in such a way so that the cosmological parameters become consistent with Planck 2018~\cite{Planck:2018vyg}.}  
In summary, Model \hyperref[M1]{I}   mimics an accelerated  dark-matter dominated 
scenario, achieved after the occurrence of the radiation phase and a 
decelerating phase where DM and baryon both coexist.

\subsection{Model II: $\Gamma = 3 \alpha H \left(\frac{\rho_{\rm c,0}}{\rho_{\rm dm}}\right)^{l}$}\label{M2}

We now proceed to the investigation of this 
model, having
the creation rate, 
\begin{equation}
\Gamma = 3 \alpha H \left(\frac{\rho_{\rm c,0}}{\rho_{\rm dm}}\right)^{l} \label{eq37},
\end{equation}
where $l$ is a   constant.\footnote{Here, ``$l$'' does not depend on any cosmological quantity. Later we will constrain it as a model parameter using various astronomical datasets.}
This generalised model extends the first model in eqn. (\ref{M1}) which corresponds to $l=1$,   by allowing for a non-linear dependence of the creation efficiency on the ambient dark matter density. 
The exponent $l$ parametrizes the sensitivity of the creation rate to the cosmological dilution of matter, thereby controlling the onset and strength of the effective negative pressure generated by the creation process. In this sense, the model represents the leading-order power-law generalization of a generic function $\Gamma(H,\rho)$, and allows for departures from the $\Lambda$CDM-like behavior driven by the earlier matter creation model in eqn. (\ref{M1}). On the other hand, we remark that as the current cosmological framework does not prefer any specific matter creation model, therefore, a generalised matter creation model is welcome until this is completely ruled out by the observational data. In other words, 

the regions of $l$ have initially been examined using the potential techniques of the dynamical systems and then it has been confronted with the observational data to understand its overall dynamics and viability.  

In this case particularly, if $ l<1/2$ then on the $\xi=0$ plane we have $H \to 
\infty,~ \rho_{\rm dm} \to \infty$ and $ \Gamma \to \infty $, while on the
$\xi=1$ plane we have $H \to 0,~ \rho_{\rm dm} \to 0$ and $ \Gamma \to 0$. 
On the other hand, for $ l>1/2$, on the  $\xi=0$ plane we have $H \to \infty,~ 
\rho_{\rm dm} \to \infty$ and $ \Gamma \to 0 $, while on the $\xi=1$ plane $H 
\to 0,~ \rho_{\rm dm} \to 0$ and $ \Gamma \to \infty$.  Finally, for $ l=1/2$, \ 
$\Gamma$ approaches to a finite value on both  $\xi=0$ and  $\xi=1$ planes. 

In this case, the conservation equation (\ref{eq20}) becomes 
\begin{equation}\label{eq40}
    \frac{d\rho_{\rm dm}}{dt} + 3 H\rho_{\rm dm} = 3\alpha H 
\left(\frac{\rho_{\rm c,0}}{\rho_{\rm dm}}\right)^l \rho_{\rm dm},
\end{equation}
which using $H = \dot{a}/a$, and imposing the 
  initial   condition    $\rho_{\rm dm}(a=1)\equiv
\rho_{\rm dm,0}$, we  extract the solution  \footnote{The solution of Eq. (\ref{eq40}) and the derivation of Eq. (\ref{m2_data}) are given in Appendix \ref{appen-B}.}
\begin{equation}
  \frac{\rho_{\rm dm}}{\rho_{c,0}} = \bigg[ \alpha + \left( \Omega_{\rm dm,0}^l 
- \alpha\right)(1+z)^{3l} \bigg]^{1/l}.
\end{equation}
As a consequence, the Hubble equation (\ref{friedmann-1A}) can be recast as 
\begin{align}\label{m2_data}
  \left(\frac{H}{H_0} \right)^2 = & \, \Bigg[  \alpha + \left(\Omega_{\rm dm,0}^l - \alpha\right)(1 + z)^{3l} \Bigg]^{1/l} \nonumber \\ 
    & + \Omega_{\rm b,0} (1 + z)^3 + \Omega_{\rm r,0}(1 + z)^4 , 
\end{align}
where again $\Omega_{\rm dm,0}+\Omega_{\rm b,0}+\Omega_{\rm r,0} =1$.  Let us note that $l= 0$  is not physically interesting, since for $l \rightarrow 0$  one   encounters a divergent  phase where $H \rightarrow \infty$.

\begin{table*}[t]    
\centering
\resizebox{1.0\textwidth}{!}{%
\begin{tabular}{|c|c|c|c|c|c|c|c|}
 \hline
   Parameter ($l$) & Fixed Points  & Eigenvalues & Stability & $\Omega_{\rm dm}$ 
& $\Omega_{\rm r}$  & $\Omega_{\rm b}$  & $q$/Acceleration(?)    \\
   \hline\hline
    &$B_0\left(1,0,1\right)$ & 
$\left(-3\alpha,-3\alpha,\frac{3}{2}\alpha\right)$, $\forall l\geq \frac{1}{2}$  
& Saddle &$1$ & $0$ & $0$  & $-\infty$; Yes \\
     
  & $B_1\left(1,0,\frac{1}{1+\sqrt[2l]{\alpha}}\right)$ & $\left(-3\alpha 
\xi_{01}^{2l},-4\alpha \xi_{01}^{2l},-3l\alpha \xi_{01}^{2l}\right)$   &   
\textbf{Stable} & $1$ & $0$ & $0$ & $-1$; Yes \\
     
  $0<l<1$ & $ B_2\left(0,1,0\right)$ & * & Unstable (Fig. \ref{fig-(ii)a})  & 
$0$ & $1$ & $0$ & $1$; No  \\
  
 & $B_3\left(0,y_c,1\right)$ & * & Unstable (Fig. \ref{fig-(ii)a})  &$0$ & $y_c$ 
& $1-y_c$ & Undetermined\\   
 & $B_4\left(x_c,0,0\right)$ & $\left(0,-1,\frac{3}{2}\right)$, $\forall l\geq 
\frac{1}{2}$ & Saddle  & $x_c$ & $0$ & $1-x_c$ & $\frac{1}{2}$; No \\

\hline 
 &$C_0\left(1,0,1\right)$ & $\left(-3\alpha,-3\alpha,\frac{3}{2}\alpha\right)$ & 
Saddle  &$1$ & $0$ & $0$  & $-\infty$; Yes \\
     
  & $C_1\left(1,0,\frac{1}{1+\sqrt[2l]{\alpha}}\right)$ &$\left(-3\alpha 
\xi_{01}^{2l},-4\alpha \xi_{01}^{2l},-3l\alpha \xi_{01}^{2l}\right)$  &   
\textbf{Stable} &  $1$ & $0$ & $0$ & $-1$; Yes \\
     
  $l>1$ & $C_2\left(0,y_c,0\right)$ & $(0,0,0)$, $\forall l\geq 2$  & Unstable  
&$0$ & $y_c$ & $1-y_c$ & Undetermined  \\
  
 & $C_3\left(x_c,0,0\right)$  & $\left(0,-x_c^{l-1},\frac{3}{2}x_c^{l-1}\right)$ 
 & Saddle if $x_c\neq 0$ & $x_c$ & $0$ & $1-x_c$ & $\frac{1}{2}$; No   \\
 
\hline 
 & $D_{0}(1,0,0)$ & $\left(-3\alpha,-3\alpha,-\frac{3\alpha}{2}\right)$, 
$\forall l\leq-\frac{1}{2}$  & \textbf{Stable } & $1$ & $0$ & $0$ & $-\infty$; 
Yes \\
 & $D_{1}\left(1,0,\frac{\sqrt[-2l]{\alpha}}{1+\sqrt[-2l]{\alpha}}\right)$ & 
$\left(-3\alpha \xi_{01}^{-2l},-4\alpha \xi_{01}^{-2l},-3l\alpha 
\xi_{01}^{-2l}\right)$  & Saddle &  $1$ & $0$ & $0$ & $-1$; Yes \\
 & $D_{2}(0,1,1)$ &$(1,1,-2)$, $\forall l\leq-\frac{1}{2}$  & Saddle   & $0$ & 
$1$ & $0$ & $1$; No \\
 $l<0 $& $D_{3}\left(0,y_c,0\right)$ & $(0,0,0)$, $\forall l \leq -\frac{1}{2} $ 
& Unstable  & $0$ & $y_c$ & $1-y_c$ & Undetermined\\
 & $D_{4}\left(x_c,0,1\right)$ & $\left(0,-1,-\frac{3}{2}\right)$, $\forall l 
\leq -\frac{1}{2}$   & \textbf{Stable} & $x_c$ & $0$ & $1-x_c$ & $\frac{1}{2}$; 
No \\
 
 \hline
 &$E_0\left(0,1,0\right)$ & $\left(2,1,1+3\alpha\right)$ & Unstable  & $0$ & $1$ 
& $0$  & $1$; No \\
     
  & $E_1\left(0,1,1\right)$ & $\left(-2,1,1+3\alpha\right)$  & Saddle &  $0$ & 
$1$ & $0$ & $1$; No \\
     
   & $E_2\left(1,0,0\right)$ & 
$\left(\frac{3}{2}(1-\alpha),-3\alpha,-1-3\alpha\right)$  & \textbf{Stable} if 
$\alpha>1$,  & $1$ & $0$ & $0$ & $\frac{1}{2}(1-3\alpha)$;   \\
   &  &   &  Saddle if $\alpha<1$  &&&& Yes if $\alpha>\frac{1}{3}$  \\
  
$l=0$ & $E_3\left(1,0,1\right)$  & 
$\left(\frac{3}{2}(\alpha-1),-3\alpha,-1-3\alpha\right)$  & \textbf{Stable} if 
$\alpha<1$, & $1$ & $0$ & $0$ & $\frac{1}{2}(1-3\alpha)$;   \\
 &  &   &  Saddle if $\alpha>1$  &&&& Yes if $\alpha>\frac{1}{3}$  \\

 &$E_4\left(0,0,0\right)$ & $\left(\frac{3}{2},-1,3\alpha\right)$ & Saddle  & 
$0$ & $0$ & $1$  & $\frac{1}{2}$; No \\

 &$E_5\left(0,0,1\right)$ & $\left(-\frac{3}{2},-1,3\alpha\right)$ & Saddle  & 
$0$ & $0$ & $1$  & $\frac{1}{2}$; No \\

 &$E_6\left(1,0,\xi_c\right)$ if $\alpha=1$ & $\left(-4,-3,0\right)$ & 
\textbf{Stable}  & $1$ & $0$ & $0$  & $-1$; Yes \\
\hline 
\end{tabular}}
\caption{Summary of the critical points, their nature (in terms of their 
stability) and the corresponding  cosmological parameters    for 
  Model II, characterized by the matter creation rate $\Gamma = 3\alpha 
\frac{H}{H_{0}}\left(\rho_{\rm c,0}/\rho_{\rm dm} \right)^l$. Here, $\xi_{01} = 
\frac{1}{1+\sqrt[2l]{\alpha}}$ and $\xi_{02} = 
\frac{\sqrt[-2l]{\alpha}}{1+\sqrt[-2l]{\alpha}}$. The asterisk (*) represents 
the failure of linear stability analysis for the critical points $B_{2}$ and 
$B_{3}$ due to singular terms 
in the Jacobian matrix, which requires numerical investigation (see text and 
Figs.\ref{fig-(ii)a}, \ref{fig-(ii)b} and  
\ref{fig-(ii)c}).  
}
\label{Table-M2}
\end{table*}

Using   (\ref{ODEs-Gen-Gamma}) and (\ref{eq37})  we obtain the following 
autonomous system 
\begin{subequations} \label{ds-(ii)}
\begin{align}
    x'=& xy+3\alpha x^{1-l}(1-x)\left(\frac{\xi}{1-\xi}\right)^{2l}, \\  
    y'=& y\left[y-1-3\alpha x^{1-l} \left(\frac{\xi}{1-\xi}\right)^{2l}\right], \\ 
    \xi'=& \frac{3}{2}\xi(1-\xi)\left[1+\frac{y}{3}-\alpha x^{1-l} 
\left(\frac{\xi}{1-\xi}\right)^{2l} \right].
\end{align}
\end{subequations} 
Moreover, the deceleration parameter (\ref{decpar})
becomes
\begin{equation}
     q= \frac{1}{2}\left[(1+y) - 3\alpha x^{1-l} 
\left(\frac{\xi}{1-\xi}\right)^{2l}  \right]. \label{q-M2}
\end{equation}
As we observe,  there can be several singularities, namely at $x=0$, $\xi=0$ 
and $\xi=1$, depending upon the values of $l$. Hence, in the following we 
examine four $l$ cases separately.

\subsubsection{Case $0<l<1$}\label{M2-(i)}  

 In this parametric domain  the system       (\ref{ds-(ii)}) 
possesses only one singularity at $\xi =1$.  Multiplying with $(1-\xi)^{2l}$ 
   yields the   autonomous system
\begin{subequations} \label{rds-(ii)a}
\begin{align}
    x'=& xy(1-\xi)^{2l}+3\alpha x^{1-l}(1-x)\xi^{2l}, \\  
    y'=& y\left[(y-1)(1-\xi)^{2l}-3\alpha x^{1-l}\xi^{2l}\right], \\ 
    \xi'=& \frac{3}{2}\xi(1-\xi)\left[\left(1+\frac{y}{3}\right)(1-\xi)^{2l}-\alpha x^{1-l} \xi^{2l} \right]. 
\end{align}
\end{subequations}
From the above dynamical  system  we acquire the following invariant manifolds: 
$x=0$, $y=0$, $\xi=0$, $\xi=1$ and $x+y=1$. Hence, the phase space 
domain $\mathbf{R}$, defined in (\ref{physical-region}), is an invariant set.

\begin{figure}[t]
    \centering
   {\includegraphics[width=0.49\textwidth]{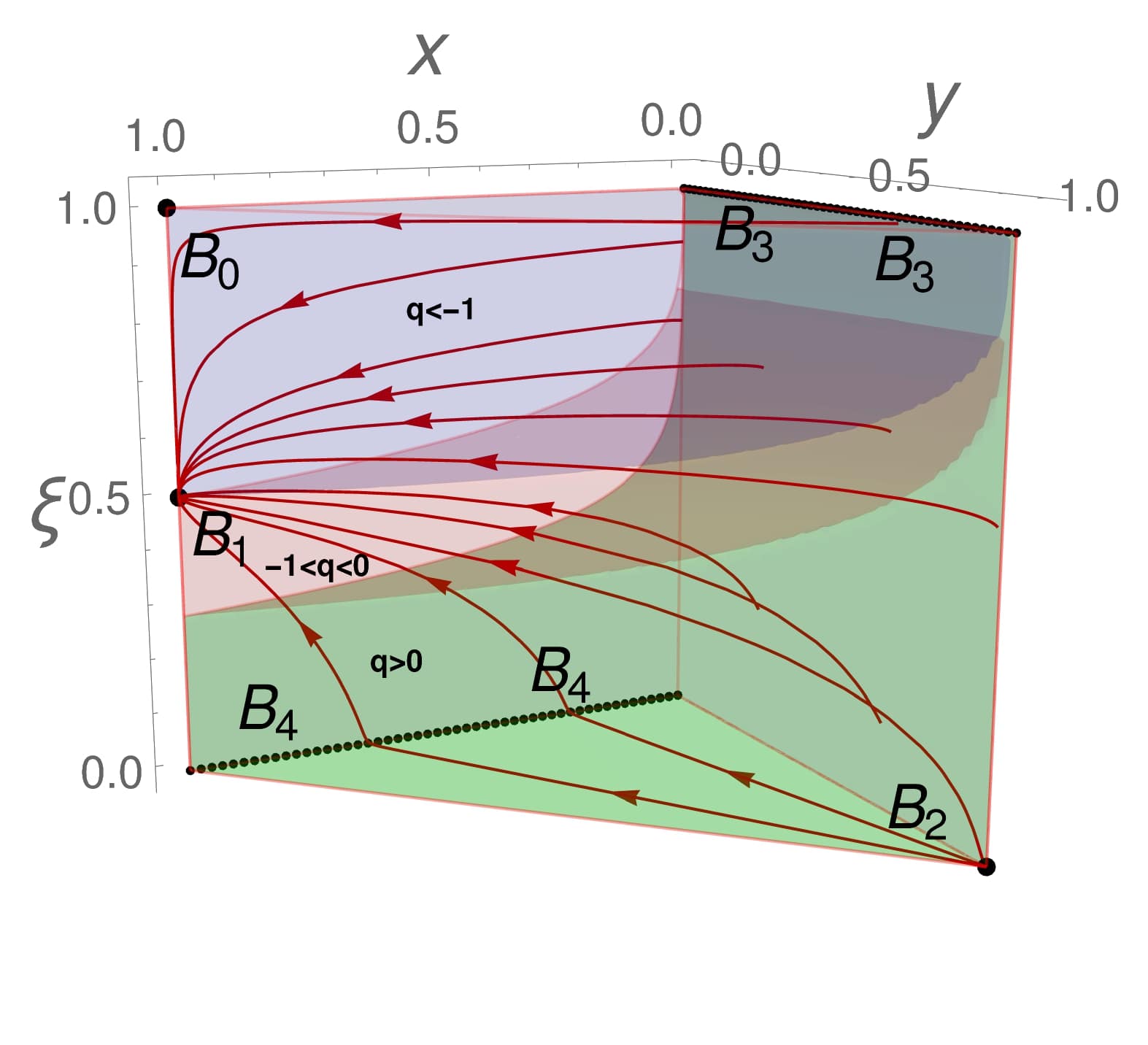}}
\caption{{\it{
Phase-space diagram for Model~\hyperref[M2-(i)]{II-1}  characterized by the 
matter creation rate $\Gamma = 3\alpha 
\frac{H}{H_{0}}\left(\rho_{\rm c,0}/\rho_{\rm dm} \right)^l$, in the case 
$0<l<1$, for $\alpha = 1$ and $l = 0.6$. The green, pink and blue 
colors   depict  cosmic deceleration, acceleration and super acceleration 
respectively.  The phase 
space possesses a completely invariant manifold, and the trajectories never 
cross the 
$x=0$ plane. We observe orbits which  emerge from the vicinity of 
$B_2$ (radiation domination plus deceleration), then  enter  into 
the accelerated  and super-accelerated regions  $B_4$, and finally are attracted 
by point  $B_1$ which is the completely matter-dominated accelerated  stable 
point.  }}}
\label{fig-(ii)a}
    \end{figure}

In Table~\ref{Table-M2}  we display the critical points, their stabilities and 
the cosmological parameters associated with them. In  particular, there are 
three isolated critical points ($B_0$, $B_1$, $B_2$) and two critical curves 
($B_3$ and $B_4$) (See Fig.~\ref{fig-(ii)a}). It is important to note that  $B_2,B_4$ and 
$B_0,B_3$ lie on the planes $\xi=0$ and $\xi=1$, respectively.
In the following,  we will systematically analyze their nature and properties. 
\begin{figure}[t]
    \centering
    \includegraphics[width=0.9\linewidth]{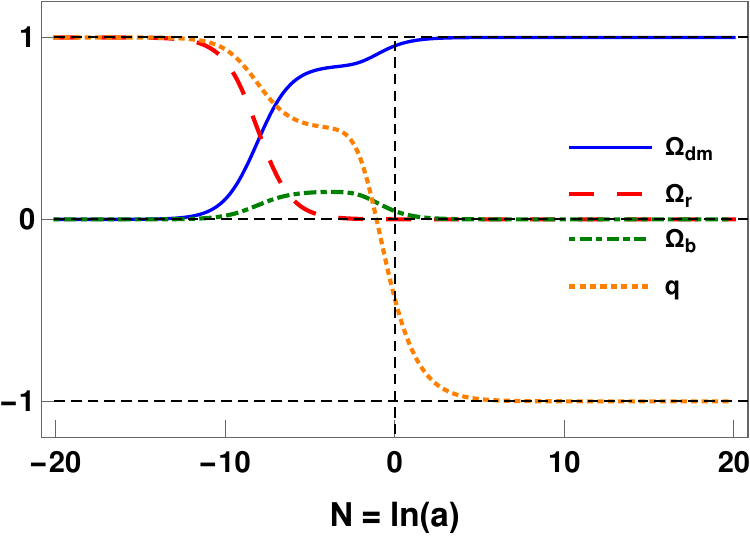}
    \caption{\textit{The evolution of the          density 
parameters $\Omega_{\rm dm}$, $\Omega_{\rm r}$, $\Omega_{\rm b}$ and
the deceleration parameter $q$, with 
respect to $\ln(a)$, for   
Model~\hyperref[M2-(i)]{II-1} in the case 
$0<l<1$, for  $\alpha=1$,  $l=0.8$, and imposing the 
initial conditions  $x(-10)=0.11$, $y(-10)=0.87$, $\xi(-10) = 0.4\times10^{-6}$. }  }
    \label{fig-M2(i)a-Om-q-N}
\end{figure}
(\texttt{i}) $\bf{B_0}$ :  
It is a saddle point, corresponding to  a  dark-matter 
dominated,  super-accelerated $(q<-1)$ solution.
 (\texttt{ii}) $ \bf{B_1}$: 
 This point is a stable point. It corresponds to a
  dark-matter dominated, accelerated de-Sitter point ($q=-1$).
 (\texttt{ii}) $\bf{B_1}$ :  
 The linear stability analysis is not applicable here due to the singular terms 
in the Jacobian matrix. However, from numerical analysis and the phase portrait 
depicted in  Fig.~\ref{fig-(ii)a}, it is clear that this is an unstable point. 
It is characterized by radiation domination and deceleration $(q>0)$.  
(\texttt{iii}) $\bf{B_2}$ :  
 The linear stability analysis is not applicable due to the singular 
terms in the Jacobian matrix. Nevertheless, from numerical analysis and the 
phase plot (Fig.~\ref{fig-(ii)a})   we deduce that it is   unstable in 
nature, characterized by the coexistence of radiation and baryon, and the
deceleration parameter remains undetermined.
(\texttt{iv}) $\bf{B_3}$ :  
This critical curve represents non-isolated saddle points 
characterized by baryon and dark matter domination, with cosmic deceleration.

Although there are some parametric limitations in the linear stability analysis, 
the phase-space diagram  can be generated for the entire range (i.e. for 
$0<l<1$).  
In   Fig.~\ref{fig-(ii)a} we observe two sets of orbits. In the 
first set, an orbit starts near the critical point $B_{2}$, indicating a 
radiation-dominated cosmic epoch. It is then affected by $B_3$, which 
corresponds to radiation and baryon domination, pulling   the orbit. 
The orbit subsequently enters the phantom super-accelerated domain as 
$\Omega_{\rm dm}$ increases. Finally, after being repelled by $B_0$, the orbit slows 
its acceleration and eventually is attracted by $B_1$,  with $ q=-1$. In the 
second set, an orbit begins near point $B_2$ and it is then repelled by the 
critical line $B_4$. Ultimately, the orbit ends at $B_1$, 
corresponding to the dark matter-dominated accelerating phase. 
 
 Hence, a typical orbit   predicts a cosmic evolution that 
begins with a radiation-dominated decelerated epoch, transits into an 
accelerated dark matter-dominated era, and then it is followed by either a dark 
matter-dominated phantom phase or a decelerating phase marked by the coexistence 
of dark matter and baryons. Ultimately, the orbit converges to a de Sitter-like 
solution at $B_1$.  

The evolution diagram, based on the phase space presented in 
Fig.~\ref{fig-(ii)a}, is given in  
Fig.~\ref{fig-M2(i)a-Om-q-N} (we consider parameter choices that do not lead to 
transient phantom crossing behavior and dominance of early acceleration right 
from the beginning of matter-radiation equality).
As we observe, our scenario 
describes a $\Lambda$CDM-like behavior, which is quite similar to the dynamics 
of Model I (see  Fig.~\ref{fig-M1-Om-q-N}).

\subsubsection{Case $l>1$}\label{M2-(ii)}

In this case the system (\ref{ds-(ii)})  simplifies  as
\begin{subequations} \label{rds-(ii)b}
\begin{align}
    x'=& x^l y(1-\xi)^{2l}+3\alpha (1-x)\xi^{2l}, \\  
    y'=& y\left[x^{l-1}(y-1)(1-\xi)^{2l}-3\alpha \xi^{2l}\right], \\ 
    \xi'=& 
\frac{3}{2}\xi(1-\xi)\left[x^{l-1}\left(1+\frac{y}{3}\right)(1-\xi)^{2l}-\alpha  
\xi^{2l} \right].
\end{align}    
\end{subequations}
The    planes $y=0$, $\xi=0$, $\xi=1$ and $x+y=1$  
constitute the invariant manifolds in the phase space. In contrast to the 
previous case, $x=0$ does not form an invariant manifold. However, since 
$x'=3\alpha \xi^{2l}>0$ at $x=0$, no trajectory from the physical domain 
$\mathbf{R}$ can escape through the $x=0$ plane. Consequently, the phase space 
remains positively invariant.

\begin{figure}[tt]
    \centering
    \includegraphics[width=0.49\textwidth]{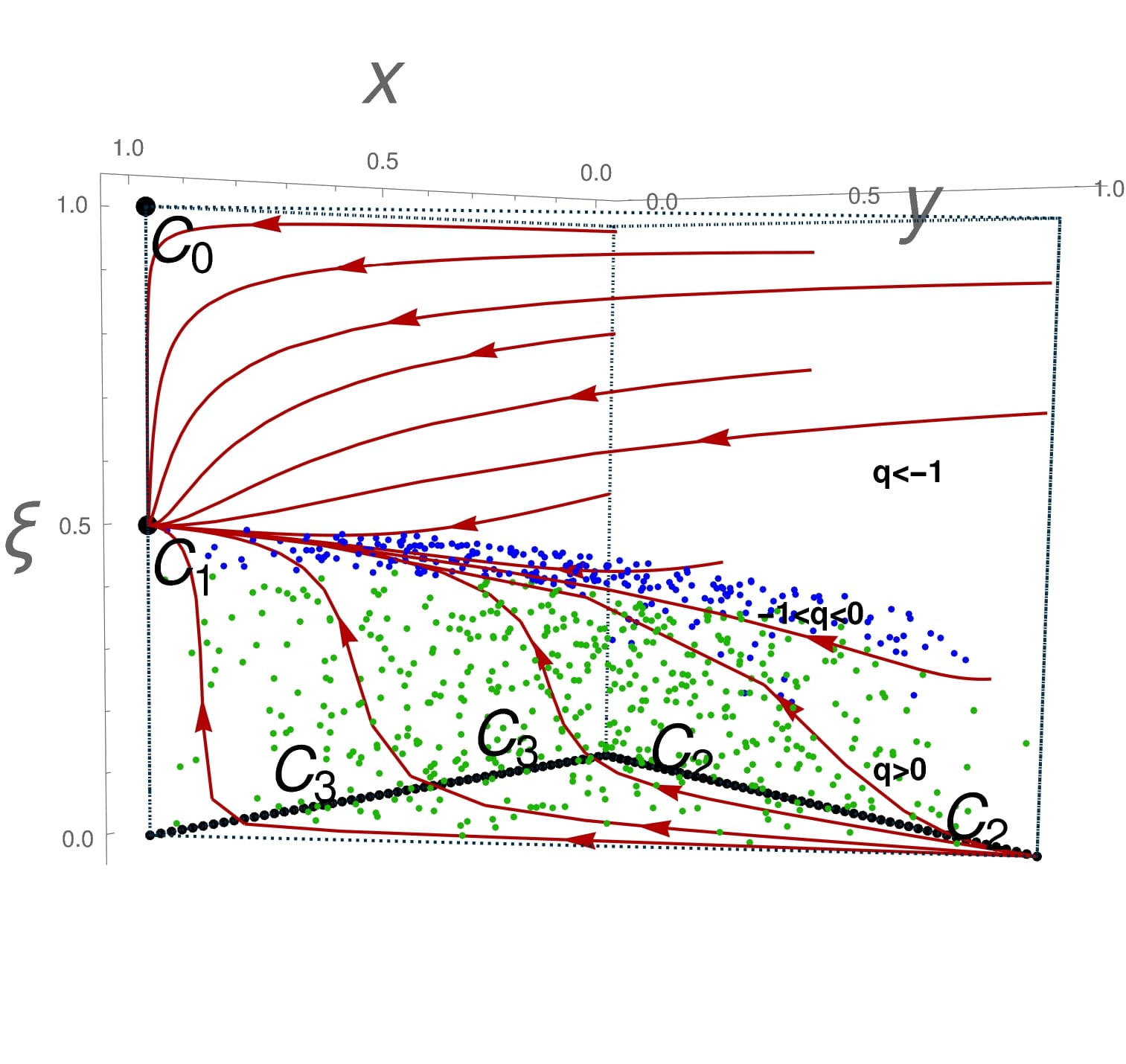}
    \caption{  {\it{
    Phase-space diagram for Model~\hyperref[M2-(i)]{II-2}  characterized by the 
matter creation rate $\Gamma = 3\alpha 
\frac{H}{H_{0}}\left(\rho_{\rm c,0}/\rho_{\rm dm} \right)^l$, in the case 
$l>1$, for $\alpha=1$ and $l=2$.    
The green dots, blue dots and undotted regions   describe deceleration, 
acceleration and super-acceleration respectively.    The incoming nature of the 
trajectories through $x=0$ plane vindicates the positively invariant nature of 
the phase space.  }}}
    \label{fig-(ii)b}
\end{figure}

In Table \ref{Table-M2}  we display the critical points  for the system 
(\ref{rds-(ii)b}), their properties and the corresponding cosmological 
parameters. In particular, we have two isolated critical points 
$(C_{0},C_{1})$ and two curves of critical points $(C_{2},C_{3})$. Let us 
examine them in detail. (See Fig.~\ref{fig-(ii)b})

    (\texttt{i}) $\bf{C_0}$ :  
This is a  saddle point, corresponding to  dark-matter dominated 
  accelerated solution.
   (\texttt{ii}) $\bf{C_1}$ :  
   This point is stable, characterized by dark matter 
domination   and acceleration. 
 (\texttt{iii}) $\bf{C_2}$ :  
   This is a curve 
   containing non-isolated critical points. 
At these critical points, the eigenvalues are all zero for $l\geq 2$, and 
numerical elaboration, as well as the phase space plot depicted in Fig. 
\ref{fig-(ii)b}, suggest  that these points are unstable, signifying a period 
where radiation and baryons coexist.  
(\texttt{iv}) $\bf{C_3}$ :  
   This critical curve of    critical points exhibits a normally 
hyperbolic character, and thus the critical points are saddle. They are 
characterized by the 
coexistence of dark and baryonic matter, and they exhibit   deceleration.

In Fig.~\ref{fig-(ii)b} we depict the phase-space trajectories. As we can see, 
 a trajectory begins in the vicinity of $C_2$, it is then 
repelled by the closure of $C_3$,   and finally it is attracted by point  
$C_1$, mimicking a de-Sitter   cosmology. 
 Hence, this scenario predicts a radiation-dominated cosmic past, followed by a 
decelerating phase where both dark matter and baryons coexist, and ultimately 
transits into an accelerated, dark matter dominated solution.
\begin{figure}[ht]
    \centering
    \includegraphics[width=0.9\linewidth]{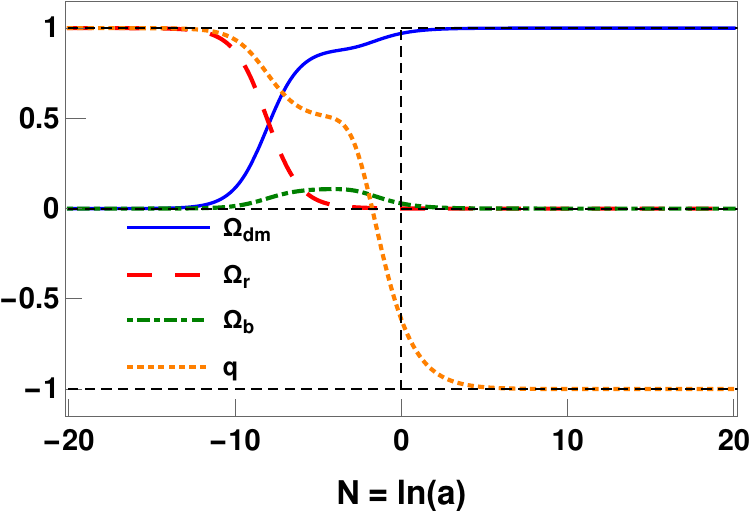}
    \caption{\textit{The evolution of the   density 
parameters $\Omega_{\rm dm}$, $\Omega_{\rm r}$, $\Omega_{\rm b}$ and
the deceleration parameter $q$, with 
respect to $\ln(a)$, for    
Model~\hyperref[M2-(ii)]{II-2} in the case $l>1$
for  $\alpha =1$, $l=1.001$, and imposing the 
initial conditions   $x(-13.5)=0.004$, $y(-13.5)=0.9955$, $\xi(-13.5) = 6\times10^{-10}$.} }
    \label{fig-M2(ii)-Om-q-N}
\end{figure}

In Fig.~\ref{fig-M2(ii)-Om-q-N}  we present the evolution   for the 
dynamical quantities. As we can see, we obtain a  $\Lambda$CDM-like 
behavior, similarly to the results obtained for Model~\hyperref[M1]{I} 
in Fig.~\ref{fig-M1-Om-q-N} and for the previous case of  
Model~\hyperref[M2-(i)]{II-1} in Fig.~\ref{fig-M2(i)a-Om-q-N}.

\subsubsection{Case $l<0$}\label{M2-(iii)}

\begin{figure}[ht]
    \centering
    \includegraphics[width=0.9\linewidth]{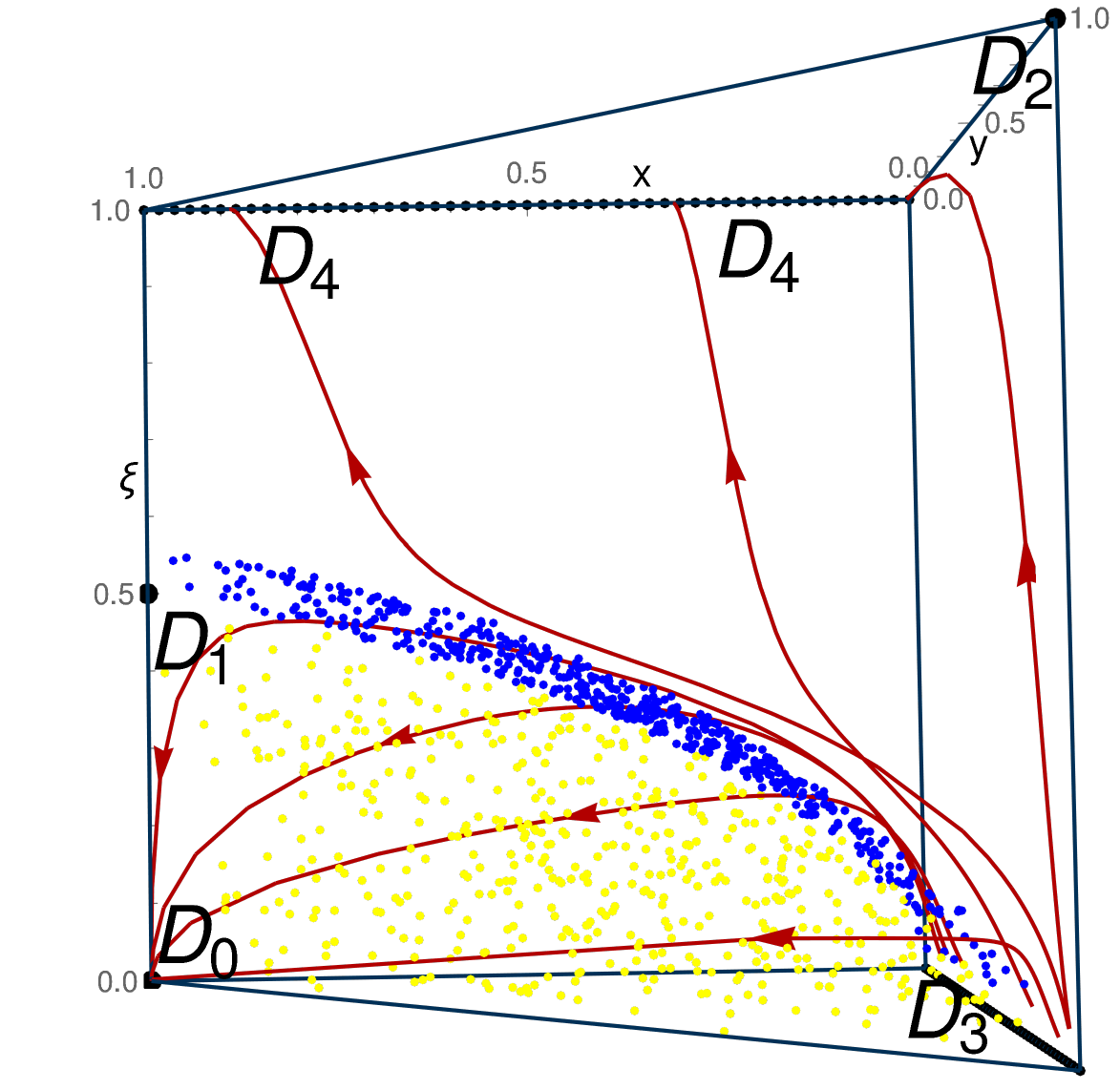}
    \caption{{\it{
      Phase-space diagram for Model~\hyperref[M2-(iii)]{II-3}   characterized 
by the 
matter creation rate $\Gamma = 3\alpha 
\frac{H}{H_{0}}\left(\rho_{\rm c,0}/\rho_{\rm dm} \right)^l$, in the case 
$l<0$, for $\alpha = 1$ and $l=-0.5$.      The white, blue-dotted and 
yellow-dotted regions denote decelerated, accelerated  and 
super-accelerated  regions respectively.  }}}
    \label{fig-(ii)c}
\end{figure}

In this case the autonomous system takes the form
\begin{subequations}\label{rds-(ii)c}
    \begin{align}
       x' =& xy\xi^{-2l} + 3\alpha x^{1-l}(1-x)(1-\xi)^{2l}, \\
       y' =& y\left[(y-1)\xi^{2l} - 3\alpha x^{1-l}(1-\xi)^{2l} \right],\\
       \xi' =& \frac{3}{2}\xi (1-\xi)\left[(1+ \frac{y}{3})\xi^{2l} - \alpha 
x^{1-l}(1-\xi)^{2l} \right],
    \end{align}
\end{subequations}
and the planes $x=0$, $y=0$, $\xi=0$, $\xi =1$ and $x+y=1 $ constitute 
invariant manifolds, making the phase space domain $\mathbf{R}$  positively 
invariant.  The critical points  and their properties are
presented in Table \ref{Table-M2}. In particular, the system exhibits  three 
isolated critical points (namely, $D_{0}$ , $D_1$ and $D_2$) and two 
curves of critical points (namely, $D_3$ and $D_4$) (See Fig.~\ref{fig-(ii)c}).  These are the following.
 (\texttt{i}) $\bf{D_0}$ :   
      This critical point is stable, dominated by dark matter, and 
super-accelerating.
 (\texttt{ii}) $\bf{D_1}$ : 
It is a saddle point (which can be shown by analytical and numerical elaboration), 
corresponding to dark-matter dominated accelerated solution.  
(\texttt{iii}) $\bf{D_2}$ : 
       It is a saddle point corresponding to radiation-dominated  
decelerating phase.
(\texttt{iv}) $\bf{D_3}$ : 
   This curve of critical points exhibits  unstable behavior (as can be shown 
by numerical elaboration and Fig. 
\ref{fig-(ii)c})), and it is characterized by   the coexistence 
of radiation and baryon. 
 (\texttt{v}) $\bf{D_4}$ : 
    The points of this critical curve are  normally  hyperbolic, exhibiting 
stable behavior, corresponding to  decelerating phase where  dark and 
baryonic matter coexist.

  In Fig.~\ref{fig-(ii)c} we depict the phase-space trajectories for this 
case. As we observe,  a set of trajectories  
start from the unstable decelerated point  $D_3$, 
next they enter into the 
accelerated phase, then they are repelled by the decelerated $D_1$ phase, and 
    finally  they are
attracted by  the dark-matter dominated   
super-accelerating solution. The fact that this scenario exhibits a   very 
early acceleration, just after the matter-radiation 
equality makes it not physically interesting, and hence we do not present the 
evolution of the various cosmological quantities. As we will see in the next 
section, observational confrontation exclude the $l<0$ region too.

\subsubsection{Case $l=0$}
\label{M2-(iv)}

\begin{figure}[t]
    \centering
\includegraphics[width=0.98\linewidth]{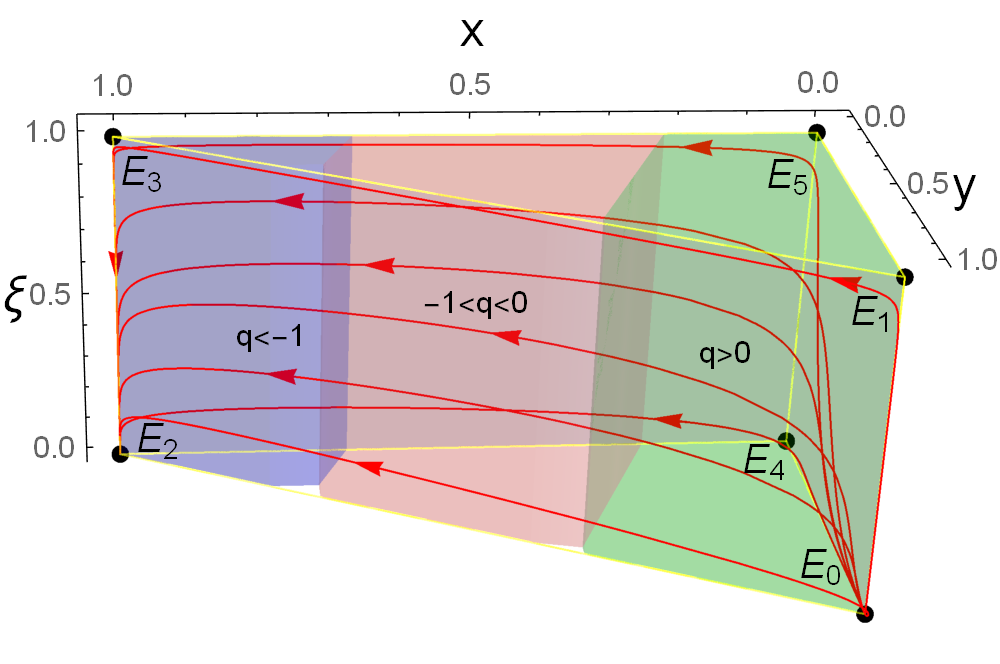}
    \caption{ {\it{   
      Phase-space diagram for Model~\hyperref[M2-(iv)]{II-4}  characterized 
by the 
matter creation rate $\Gamma = 3\alpha 
\frac{H}{H_{0}}\left(\rho_{\rm c,0}/\rho_{\rm dm} \right)^l$, in the case 
$l=0$, for  $\alpha = 1.5$.      
    The green, pink and blue shaded regions highlight decelerated, 
accelerated, and super-accelerated   regions respectively.   }}}
    \label{fig-(ii)d}
\end{figure}

In the case $l=0$ the autonomous system simplifies significantly, since 
$\Gamma=3\alpha H$, namely it becomes
\begin{subequations} \label{rds-(ii)d}
\begin{align}
    x'=& xy+3\alpha x(1-x), \\  
    y'=& y\left(y-1-3\alpha x\right), \\ 
    \xi'=& \frac{3}{2}\xi(1-\xi)\left(1+\frac{y}{3}-\alpha x \right).
\end{align}
\end{subequations}
The reduced system possesses five invariant manifolds, namely $x=0$, $y=0$, 
$\xi=0$, $\xi=1$ and $x+y=1$, ensuring that the phase space domain $\mathbf{R}$ 
again remains positively invariant. In   Table \ref{Table-M2}  we present  
  the critical points and their properties. The system exhibits  six 
critical points $\left(E_0,E_1,E_2,E_3,E_4,E_5\right)$, and one curve 
of critical points $\left(E_6\right)$ (See Fig.~\ref{fig-(ii)d}). These are the following.
\begin{figure}[t]
    \centering
    \includegraphics[width=0.9\linewidth]{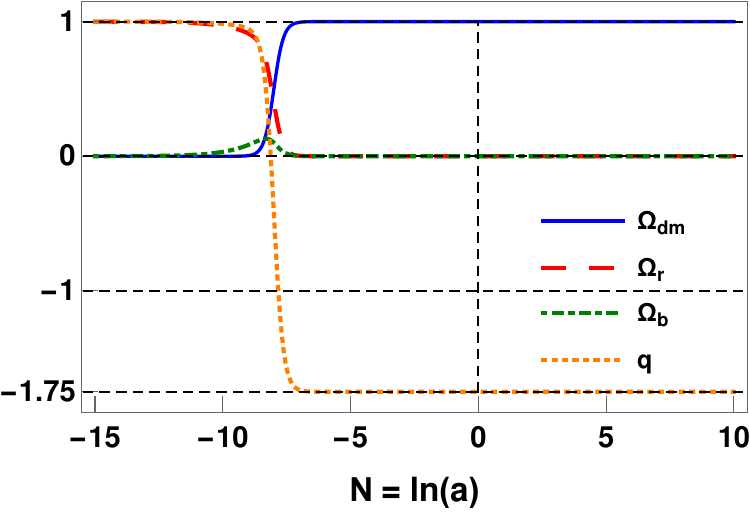}
    \caption{ \textit{The evolution of the          density 
parameters $\Omega_{\rm dm}$, $\Omega_{\rm r}$, $\Omega_{\rm b}$ and
the deceleration parameter $q$, with 
respect to $\ln(a)$, for    
 Model~\hyperref[M2-(iv)]{II-4} in the case $l=0$,
$\alpha =1.5$, and imposing the 
initial conditions $x(-8)=0.53$, $y(-8)=0.44$, $\xi(-8) = 0.8\times10^{-5}$. }}
    \label{fig-M2(iv)-Om-q-N}
\end{figure}
(\texttt{i}) $\bf{E_0}$ : It is an unstable point, corresponding to 
radiation-dominated, decelerating solution.
 (\texttt{ii}) $\bf{E_1}$ :
It is a saddle point, corresponding to radiation-dominated, decelerating 
solution. 
 (\texttt{iii}) $\bf{E_2}$ :
 It is a stable point for $\alpha>1$, corresponding to   dark-matter dominated  
accelerated solution.
 (\texttt{iv}) $\bf{E_3}$ :
  It is a stable point for $\alpha<1$ corresponding to   dark-matter dominated  
accelerated solution for $\alpha>\frac{1}{3}$.
 (\texttt{v}) $\bf{E_4}$ :
  It is a saddle point, corresponding to  baryon-dominated, decelerating 
solution.
 (\texttt{vi}) $\bf{E_5}$ :
  It is a saddle point, corresponding to baryon-dominated, decelerating 
solution.
 (\texttt{vii}) $\bf{E_6}$ : The points of this curve of critical points (containing 
also the critical points $E_2$ and $E_3$) are normally hyperbolic, and
numerical elaboration shows that they behave as stable points. They correspond to
 dark-matter dominated  
accelerated solution.

\begin{figure}[t]
    \centering
    \includegraphics[width=0.9\linewidth]{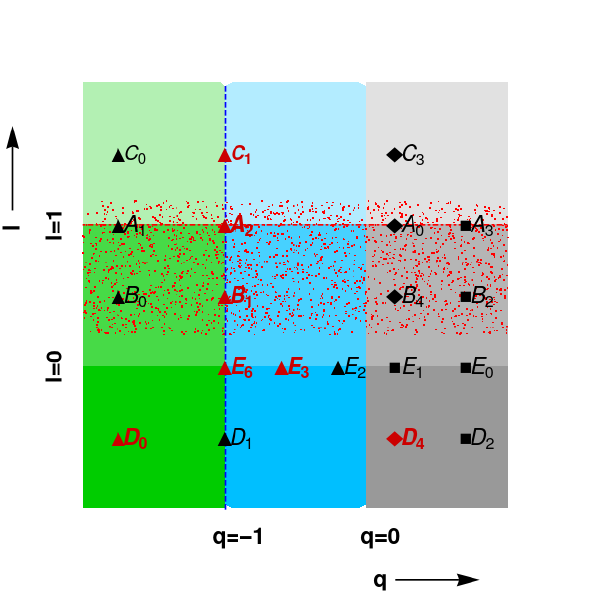}
    \caption{{\it{Summary plot of the obtained results.    
    The horizontal and vertical directions are 
the deceleration parameter $q$ and the model parameter $l$,
respectively. The vertical green, blue and grey colors distinguish among the 
different acceleration regions, whereas the different parametric regions, 
$-\infty \leq l \leq 0$ , $0\leq l \leq 1$ and $1 \leq l \leq \infty$ have been 
distinguished by different opaques.   The ``Triangle'', ``Diamond'' and 
``Square'' shapes represent $\Omega_{\rm dm}=1$, $\Omega_{\rm dm}+\Omega_{\rm 
b}=1$ and $\Omega_{\rm r}=1$ respectively. The `red-colored'' critical points 
are the stable ones, while  the ``black-colored'' critical points represent 
unstable or saddle points. Finally, the red-dusted region marks the 
observationally viable limits of $l$  obtained through a detailed data analysis 
presented in the next section.
    }}}
    \label{fig-summary}
\end{figure}

 In Fig.~\ref{fig-(ii)d} we draw the phase-space portrait of the $l=0$ 
sub-case.  Here the class of trajectories start from a 
radiation-dominated decelerated point ($E_0$), then they   are repelled by 
decelerated radiation and baryon-dominated saddle points ($E_1$, $E_4$ and 
$E_5$), then, with increasing matter domination, the trajectories 
enter the accelerated region, and finally they result to a 
super-accelerated matter dominated solution.  
 Furthermore, in Fig.~\ref{fig-M2(iv)-Om-q-N} we show the evolution of the 
density parameters and the deceleration parameter. The fact that $q$ becomes 
negative very early, i.e. the system exhibits 
  an early acceleration like Model~\hyperref[M2-(iii)]{II-3}  and 
 implies that it is not physically interesting, a 
result that is going to be verified by observational confrontation, too.

We close this section by presenting a summary plot of all the 
dynamical-analysis results of both Model I and Model II, in  
Fig.~\ref{fig-summary}.

\section{Observational datasets and methodology}
\label{sec-data}

In this section we describe the observational  datasets and the methodology used 
to constrain the proposed matter creation models. In what follows we outline the 
datasets.

\begin{enumerate}
    \item Hubble data from Cosmic Chronometers ({\bf CC}): We use 34 
measurements of the Hubble parameter at different redshifts using the CC 
approach, displayed in   Table~\ref{table_CC}. The $\chi^2$ is defined as, 
$\chi^2 = \sum_{\rm i} \left(\frac{H_{\rm CC,i} - H_{\rm model, i}}{\sigma_{\rm i}}\right)^2$, where `i' denotes the redshift values.

\begin{table}[ht]
\begin{center}
\resizebox{0.47\textwidth}{!}{%
    \tiny 
\renewcommand{\arraystretch}{1.0} 
\begin{tabular}{l c c r|l c c r}
\multicolumn{8}{c}{}\\
\hline 
$z$ & $H(z)$  & $\sigma_{H(z)}$  & Ref. ~& $~z$ & $H(z)$ & $\sigma_{H(z)}$  & 
Ref.\\
\hline

0.07 	&   69.0   &  19.6	&  \cite{Zhang:2012mp}~ & ~0.48 	&   97.0   &  
62.0	&  \cite{Stern:2009ep} \\
0.09 	&   69.0   &  12.0	&  \cite{Jimenez:2003iv}~ & ~0.5929 	&   104.0   
&  13.0	&  \cite{Moresco:2012by} \\
0.12 	&   68.6   &  26.2	&  \cite{Zhang:2012mp}~ & ~0.6797 	&   92.0   &  
8.0	&  \cite{Moresco:2012by} \\
0.17 	&   83.0   &  8.0	&  \cite{Simon:2004tf}~ & ~0.75 	&   98.8   &  
33.6	&  \cite{Borghi:2021rft} \\
0.1791 	&   75.0   &  4.0	&  \cite{Moresco:2012by}~ & ~0.7812 	&   105.0   
&  12.0	&  \cite{Moresco:2012by} \\
0.1993 	&   75.0   &  5.0	&  \cite{Moresco:2012by}~ & ~0.80 	&   113.1   &  
25.22	&  \cite{Jiao:2022aep} \\
0.20 	&   72.9   &  29.6	&  \cite{Zhang:2012mp}~ & ~0.8754 	&   125.0   &  
17.0	&  \cite{Moresco:2012by} \\
0.27 	&   77.0   &  14.0	&  \cite{Simon:2004tf}~ & ~0.88 	&   90.0   &  
40.0	&  \cite{Stern:2009ep} \\
0.28 	&   88.8   &  36.6	&  \cite{Zhang:2012mp}~ & ~0.90 	&   117.0   &  
23.0	&  \cite{Simon:2004tf} \\
0.3519 	&   83.0   &  14.0	&  \cite{Moresco:2012by}~ & ~1.037 	&   154.0   &  
20.0	&  \cite{Moresco:2012by} \\
0.3802 	&   83.0   &  13.5	&  \cite{Moresco:2016mzx}~ & ~1.26 	&   135.0   &  
65.0	&  \cite{Tomasetti:2023kek} \\
0.4 	&   95.0   &  17.0	&  \cite{Simon:2004tf}~ & ~1.3 	&   168.0   &  17.0	
&  \cite{Simon:2004tf} \\
0.4004 	&   77.0   &  10.2	&  \cite{Moresco:2016mzx}~ & ~1.363 	&   160.0   
&  33.6	&  \cite{Moresco:2015cya} \\
0.4247 	&   87.1   &  11.2	&  \cite{Moresco:2016mzx}~ & ~1.43 	&   177.0   &  
18.0	&  \cite{Simon:2004tf} \\
0.44497 	&   92.8   &  12.9	&  \cite{Moresco:2016mzx}~ & ~1.53 	&   140.0   
&  14.0	&  \cite{Simon:2004tf} \\
0.47 	&   89.0   &  49.6	&  \cite{Ratsimbazafy:2017vga}~ & ~1.75 	&   
202.0   &  40.0	&  \cite{Simon:2004tf} \\
0.4783 	&   80.9   &  9.0	&  \cite{Moresco:2016mzx}~ & ~1.965 	&   186.5   
&  50.4	&  \cite{Moresco:2015cya} \\
\hline
\end{tabular}
}

\end{center}

\caption{The 34 measurements of the Hubble parameter at different redshifts 
using the Cosmic Chronometers  
approach, used in this work, obtained from the mentioned references.}

\label{table_CC}
\end{table}

\item Type Ia Supernovae ({\bf SNIa}): We   use  three distinct samples of SNIa 
data.  
\begin{table*}[t]
\centering
    \begin{tabular}{cccccccc}
    \hline
       & &  Model I & &  \\
       \hline
          Dataset & $H_0$ (km/s/Mpc)  & $\alpha$ &$\mathcal{M}$ \\
          \hline
        
          CC+PP & $67.70\pm 1.70$  &$0.670\pm0.018$ & $23.8071\pm 0.0068$   \\
          CC+PP+D-DR1 & $68.52\pm 0.68$ &$0.690 \pm 0.011$ & $23.8005\pm 0.0054$ 
 \\
         
          CC+PP+D-DR2  &$68.73\pm 0.45$  &$0.6957\pm 0.0078$ &$23.7988\pm0.0046$ 
  \\
         CC+D5 & $69.52\pm 0.33$ & $0.659 \pm 0.015$ & $23.7804 \pm 0.0070$ \\
         
         CC+D5+D-DR1 &  $70.43 \pm 0.23$ & $0.7101 \pm 0.0073$& $23.7798\pm 0.0056$ \\
         
          CC+D5+D-DR2 & $70.48\pm 0.19$ &$0.7178\pm 0.0048$ & $23.7786\pm 0.0044$ \\
         CC+U3 & $68.50 \pm 1.60$ & $0.661 \pm 0.023$ & $23.8125\pm 0.0065$  \\
           CC+U3+D-DR1 & $68.69 \pm 0.72$ & $0.693 \pm 0.012$ & $23.8010\pm 0.0052$  \\
          CC+U3+D-DR2 &$68.83\pm 0.46$ & $0.6973\pm 0.0081$& $23.7987\pm 0.0045$ \\
          \hline
         
    \end{tabular} 
    
    \caption{68\% confidence-level constraints on the model parameters 
corresponding to Model I. Here PP, D5, U3, D-DR1 and D-DR2 respectively refer 
to PantheonPlus, DESY5, Union3, DESI-BAO DR1 and DESI-BAO DR2.  }
    \label{Tab-data-M1}    
\end{table*}
\begin{table*}[t]
\centering
    \begin{tabular}{cccccccc}
    \hline
        & & Model II & & \\
       \hline
          Dataset & $H_0$ (km/s/Mpc) & $l$ &$\alpha$ & $\mathcal{M}$\\
          \hline
        
          CC+PP  & $68.10\pm 1.80$ &$0.830^{+0.220}_{-0.170}$ & 
$0.653^{+0.030}_{-0.024}$ &$23.8103\pm 0.0080$ \\
          CC+PP+D-DR1 & $68.16\pm 0.72$& $0.875\pm 
0.088$&$0.663^{+0.024}_{-0.021}$ & $23.8080\pm0.0073$ \\
         
          CC+PP+D-DR2  & $67.96\pm 0.58$ &$0.862\pm 0.068$ & $0.659\pm0.021$ 
&$23.8089\pm 0.0069$ \\
         CC+D5 & $ 69.18\pm0.40$& $0.670\pm 0.31$ & $0.619^{+0.033}_{-0.029}$ & 
$23.7908\pm 0.0088$ \\
         
         CC+D5+D-DR1 &  $69.71^{+0.34}_{-0.30}$ & $0.777^{+0.076}_{-0.057}$ & 
$0.662^{+0.020}_{-0.015}$ & $23.7885\pm 0.0072$ \\
         
          CC+D5+D-DR2 & $70.07\pm 0.31$ &$0.890\pm 0.062$ & $0.693 \pm 0.015$ 
&$23.7815\pm 0.0068$ \\
         CC+U3  & $68.40\pm 1.60$ & $0.650^{+0.390}_{-0.380}$ & 
$0.607^{+0.051}_{-0.043}$ & $23.8080\pm 0.0062$ \\
           CC+U3+D-DR1 & $67.75\pm 0.89$ & $0.817^{+0.098}_{-0.110}$ & $0.643\pm 
0.032$  & $23.8098\pm 0.0048$ \\
          CC+U3+D-DR2 & $67.67\pm0.72 $&$0.821\pm 0.082$ &$0.645\pm 0.027$  & 
$23.8022\pm 0.0040$\\
          \hline
         
    \end{tabular} 
    \caption{68\% confidence-level constraints on the model parameters 
corresponding to Model II. Here PP, D5, U3, D-DR1 and D-DR2 respectively refer 
to PantheonPlus, DESY5, Union3, DESI-BAO DR1 and DESI-BAO DR2. 
}
    \label{Tab-data-M2}

\end{table*}
\begin{table*}[t]
\scriptsize
\centering
\resizebox{0.7\textwidth}{!}{%
\begin{tabular}{cccccccc}
\hline 
Data & Models & $\chi_{\rm min}^2$ & $\Delta\chi_{\rm min}^2$ & AIC & $\Delta$AIC & BIC & $\Delta$BIC \\[1ex]
\hline 

\raisebox{-1ex}{CC+PP} & $M_1$ & 1418.36007 & \raisebox{-1ex}{$0.00292$} & 1424.36007 & \raisebox{-1ex}{$0.00292$} & 1440.53801 & \raisebox{-1ex}{$0.00292$} \\
                         & $\Lambda$CDM & 1418.36299 & & 1424.36299 & & 1440.54093 & \\

\hline

\raisebox{-1ex}{CC+PP+D-DR1} & $M_1$ & 1433.58257 & \raisebox{-1ex}{$0.00206$} & 1439.58257 & \raisebox{-1ex}{$0.00206$} & 1455.78260 & \raisebox{-1ex}{$0.00205$} \\
                         & $\Lambda$CDM & 1433.58463 & & 1439.58463 & & 1455.78465 & \\

\hline

\raisebox{-1ex}{CC+PP+D-DR2} & $M_1$ & 1431.47123 & \raisebox{-1ex}{$-0.0036$} & 1437.47123 & \raisebox{-1ex}{$-0.0036$} & 1453.67310 & \raisebox{-1ex}{$-0.00361$} \\
                         & $\Lambda$CDM & 1431.46763 & & 1437.46763 & & 1453.66949 & \\

\hline

\raisebox{-1ex}{CC+D5} & $M_1$ & 1656.93580 & \raisebox{-1ex}{$0.00926$} & 1660.93580 & \raisebox{-1ex}{$0.00926$} & 1671.91350 & \raisebox{-1ex}{$0.00927$} \\
                         & $\Lambda$CDM & 1656.94506 & & 1660.94506 & & 1671.92277 & \\

\hline

\raisebox{-1ex}{CC+D5+D-DR1} & $M_1$ & 1689.29975 & \raisebox{-1ex}{$-0.01989$} & 1693.29975 & \raisebox{-1ex}{$-0.01989$} & 1704.29083 & \raisebox{-1ex}{$-0.01988$} \\
                         & $\Lambda$CDM & 1689.27986 & & 1693.27986 & & 1704.27095 & \\

\hline

\raisebox{-1ex}{CC+D5+D-DR2} & $M_1$ & 1698.01771 & \raisebox{-1ex}{$0.00532$} & 1702.01771 & \raisebox{-1ex}{$0.00532$} & 1713.00990 & \raisebox{-1ex}{$0.00533$} \\
                         & $\Lambda$CDM & 1698.02303 & & 1702.02303 & & 1713.01523 & \\

\hline

\raisebox{-1ex}{CC+U3} & $M_1$ & 41.55208 & \raisebox{-1ex}{$0.00207$} & 45.55208 & \raisebox{-1ex}{$0.00207$} & 49.60278 & \raisebox{-1ex}{$0.00208$} \\
                         & $\Lambda$CDM & 41.55415 & & 45.55415 & & 49.60486 & \\

\hline

\raisebox{-1ex}{CC+U3+D-DR1} & $M_1$ & 57.50710 & \raisebox{-1ex}{$0.00007$} & 61.50710 & \raisebox{-1ex}{$0.00007$} & 65.94611 & \raisebox{-1ex}{$0.00006$} \\
                         & $\Lambda$CDM & 57.50703 & & 61.50703 & & 65.94605 & \\

\hline

\raisebox{-1ex}{CC+U3+D-DR2} & $M_1$ & 55.24688 & \raisebox{-1ex}{$0.00154$} & 59.24688 & \raisebox{-1ex}{$0.00154$} & 63.71509 & \raisebox{-1ex}{$0.00155$} \\
                         & $\Lambda$CDM & 55.24842 & & 59.24842 & & 63.71664 & \\

\hline 
\end{tabular}}
\caption{Comparison between Model I and $\Lambda$CDM paradigm, through the 
statistical parameters $\chi_{\rm min}^2$, AIC, and BIC.}
\label{tab-m1-lcdm-chi2}
\end{table*}

\begin{table*}[t]
    \centering
    \resizebox{0.7\textwidth}{!}{%
    \begin{tabular}{ccccccccccccc}
    \hline 
        Data & Models & $\chi
        _{\rm min}^2$ & $\Delta\chi
        _{\rm min}^2$ & $\rm{AIC}$ & $\Delta \rm{AIC}$ & $\rm{BIC}$ & $\Delta \rm{BIC}$ \\[1ex]
        \hline

        \raisebox{-1ex}{CC+PP} & $M_2$ &1418.15688 & \raisebox{-1ex}{$0.20611$}& 1426.15688&\raisebox{-1ex}{$-1.79389$}  & 1447.72747 & \raisebox{-1ex}{$-7.18654$} \\
        
         & $\Lambda$CDM & 1418.36299 & & 1424.36299 & & 1440.54093 &    \\ 
         
          \hline 
         
         \raisebox{-1ex}{CC+PP+D-DR1}& $M_2$ &1431.33889 & \raisebox{-1ex}{$2.24574$}& 1439.33889& \raisebox{-1ex}{$0.24574$}& 1460.93892 & \raisebox{-1ex}{$-5.15427$} \\
        
            & $\Lambda$CDM & 1433.58463 & & 1439.58463 & & 1455.78465 &   \\

            \hline 
           \raisebox{-1ex}{CC+PP+D-DR2} & $M_2$ & 1427.46062 &\raisebox{-1ex}{$4.00701$}& 1435.46062 & \raisebox{-1ex}{$2.00701$}& 1457.06310  & \raisebox{-1ex}{$-3.39361$} \\
         & $\Lambda$CDM & 1431.46763 & & 1437.46763 & & 1453.66949&    \\
           \hline 
         \raisebox{-1ex}{CC+D5}  & $M_2$ &1654.85185 & \raisebox{-1ex}{$2.09321$}& 1660.85185 & \raisebox{-1ex}{$ 0.09321$}& 1677.31841 & \raisebox{-1ex}{$-5.39564$} \\
         & $\Lambda$CDM & 1656.94506 & & 1660.94506 & &  1671.92277 &   \\

           \hline 
         \raisebox{-1ex}{CC+D5+D-DR1} & $M_2$ & 1682.48617&\raisebox{-1ex}{$6.79369$}  & 1688.48617&\raisebox{-1ex}{$4.79369$} & 1704.97280 & \raisebox{-1ex}{$-0.70185$} \\
            & $\Lambda$CDM & 1689.27986 & &1693.27986 & & 1704.27095 &  \\

             \hline 
            \raisebox{-1ex}{CC+D5+D-DR2}& $M_2$ &1695.14234 & \raisebox{-1ex}{$2.88069$}&1701.14234 & \raisebox{-1ex}{$0.88069$}& 1717.63063&  \raisebox{-1ex}{$-4.6154$} \\
            & $\Lambda$CDM & 1698.02303 & & 1702.02303 & & 1713.01523 &   \\
               \hline     
           \raisebox{-1ex}{CC+U3} & $M_2$ &39.87716 &\raisebox{-1ex}{$1.67699$} & 45.87716 & \raisebox{-1ex}{$ -0.32301$}& 51.95322 &  \raisebox{-1ex}{$-2.34836$}   \\
              & $\Lambda$CDM &41.55415 & &45.55415 & &49.60486  &  \\

              \hline  
             \raisebox{-1ex}{CC+U3+D-DR1} & $M_2$ &54.56990 & \raisebox{-1ex}{$2.93713$}& 60.56990&\raisebox{-1ex}{$0.93713$} & 67.22842 & \raisebox{-1ex}{$-1.28237$}  \\
              & $\Lambda$CDM & 57.50703 & &61.50703 & & 65.94605  &   \\
                 \hline

               \raisebox{-1ex}{CC+U3+D-DR2}& $M_2$ & 50.73210 & \raisebox{-1ex}{$4.51632$}& 56.73210& \raisebox{-1ex}{$2.51632$}& 63.43442 &\raisebox{-1ex}{$0.28222$}   \\
                & $\Lambda$CDM & 55.24842& & 59.24842& & 63.71664 &    \\
                \hline

    \end{tabular}}
    \caption{Comparison between Model II and $\Lambda$CDM paradigm, through the 
statistical parameters $\chi_{\rm min}^2$, AIC and BIC. }
    \label{tab-m2-lcdm-chi2}
\end{table*}

\begin{figure}[t]
    \centering
    \subfigure[ Model 
I]{\includegraphics[width=0.42\textwidth]{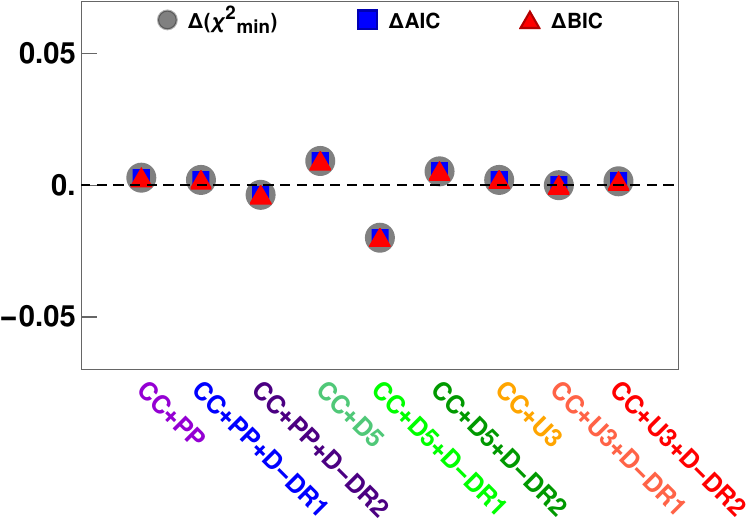}}\label{fig-M1-aic-bic}
\\
    \subfigure[ Model 
II]{\includegraphics[width=0.41\textwidth]{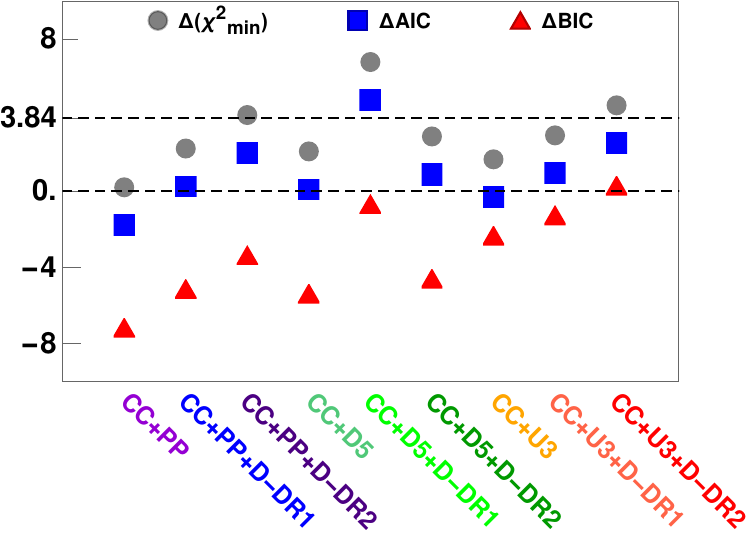}}\label{fig-M2-aic-bic
}
    \caption{{\it{Summary of  the 
values of $ \Delta \chi_{\rm min}^2$, $\Delta \rm{AIC}$ and $\Delta \rm{BIC}$  
for  Model I and Model II, considering all     dataset  combination (see text 
and Tables \ref{tab-m1-lcdm-chi2} and \ref{tab-m2-lcdm-chi2}).}
    \label{fig-aic-bic}}}
\end{figure}  

\begin{enumerate}

\item \textbf{PantheonPlus (PP)}:  In PP dataset we   exclude  the 
observational results from SH0ES~\cite{Brout:2022vxf,Riess:2021jrx}, by 
eliminating $z<0.01$ redshift-dependent results and the absolute magnitude 
calibration made by SH0ES. The apparent and absolute magnitude are related by  
$m_{\rm B}- M_{\rm B} = 5 \ {\rm log_{10}}\left(\frac{d_{\rm L}}{1\ \rm Mpc}\right) +25$, where  $d_{\rm L} = 
\frac{c(1+z)}{H_0}\int_{0}^{z} \frac{dz'}{E(z')}$ is the luminosity distance, 
and    $m_{\rm B} = 5{\rm 
log_{10}}\left((1+z)\int_{0}^{z} \frac{dz'}{E(z')}\right) + \mathcal{M}$ 
is the apparent luminosity magnitude. $\mathcal{M}$ is defined by, $\mathcal{M}= 25 +M_{\rm b} + \log_{10} (\frac{c / H_0}{1 \mathrm{Mpc}}) $, where $M_{\rm B}$ is the absolute luminosity magnitude and $c$ is the speed of light. 
We use $m_{\rm B}$ instead of the distance moduli $\mu$ ($\mu = m_{\rm B} - M_{\rm B}$) to construct the likelihood function because it preserves the intrinsic degeneracy between $M_B$ and $H_0$, preventing an artificial calibration that would otherwise spuriously fix $H_0$ when combining supernova datasets like Pantheon Plus, DES~Y5 and Union3. The likelihood function is Gaussian, $\mathcal{L} = e^{-\frac{\chi^2}{2}}$. We construct the $\chi^2$ for the PP data with the difference between apparent luminosity magnitude from the data and the model. $\Delta m_{\rm B} = m_{\rm B, data} - m_{\rm B, model}$. 
Hence, $\chi^2 = \Delta m_{\rm B,i}^{T} \mathcal{C}_{\rm ij}^{-1} \Delta m_{\rm B,j}$. Where $\mathcal{C}$ is the covariance matrix consisting of both statistical and systematic errors of the PP data.

\item  \textbf{DES Year 5}: 
The Dark Energy Survey (DES) 5-year  
analysis~\cite{DES:2024jxu,DES:2024hip,DES:2024upw}, utilizing its full 
dataset\footnote{\url{https://github.com/CobayaSampler/sntextunderscore data/tree/master/DESY5}}  of 1829 supernova (1635 number of SNIa from the 5 year 
sample and 94 number of SNIa from low-red-shift external sample), represents a 
significant milestone in the study of the universe's large-scale structure and 
the nature of dark energy. We use the filtered dataset for the 
redshift $z > 0.01$, which   consists of 1754 number of distance moduli for 
SNIa. 
The $\chi^2$ is defined as, $\chi_{\rm DES}^2 = \Delta m_{\rm B,i}^{\rm T}\mathcal{C}_{\rm ij}^{-1}\Delta m_{\rm B,i}$, where $m_{\rm B}$ denotes apparent magnitude and $\Delta m_{\rm B, i} = m_{\rm data, i} - m_{\rm model, i}$ and $\mathcal{C}$ being the covariance matrix for the DES Year 5 data.

\item \textbf{Union3}: 
Though the full dataset has not been come up yet, the Union3 (U3) supernova 
dataset,\footnote{\url{https://github.com/CobayaSampler/sn\_data/tree/master/Union3}} incorporating the 22-binned data~\cite{Rubin:2023ovl} 
from 2087 SNIa, serves as a critical tool for constraining cosmological models 
and probing the expansion history of the universe. 
For Union3  $\chi^2$ is expressed as, $\chi_{\rm union3}^2 = \Delta m_{\rm B, i}^{\rm T}\mathcal{C}_{\rm ij}^{-1}\Delta m_{\rm B, j}$, where $m_{\rm B}$ is the apparent magnitude and $\Delta m_{\rm B, i} = m_{\rm B, data, i} - m_{\rm B, model, i}$ and  $\mathcal{C}$ being the covariance matrix for the Union3 data. 

\end{enumerate}

\item \textbf{DESI-BAO}:    DESI (Dark Energy Spectroscopic Instrument) 
BAO (Baryon Acoustic 
Oscillation)~\cite{DESI:2024uvr,DESI:2024lzq,Cortes:2024lgw,DESI:2024mwx} 
analysis utilizes binned data to derive precise measurements of the angular 
diameter distance, Hubble distance, and velocity-averaged distance. By mapping 
the large-scale structure   with  high precision across 
millions of galaxies and quasars, DESI achieves high-resolution constraints on 
the BAO scale. The binning methodology enables refined measurements of 
cosmological distances as a function of redshift, minimizing statistical noise 
and systematic biases. 

The co-moving distance is
$D_{\rm M} = c\int_{0}^{z} \frac{dz'}{H(z')}$, while
the Hubble distance is $D_{\rm H} = {c}/{H(z)}$, and 
the volume-averaged distance is 
$D_{\rm V} =\sqrt[3]{z  D_{\rm M}^{2} D_{\rm H} }$.  
DESI has 7 redshift bins from the wide  spectroscopical survey in which the 
mission has calculated a set of the above    quantities over 
the sound horizon radius at baryon drag epoch, e.g.  ${D_{\rm M}}/{r_{\rm d}}$, ${D_{\rm H}}/{r_{\rm d}}$ 
and ${D_{\rm V}}/{r_{\rm d}}$. Hence, we  calculate the model values 
of these quantities, then  we find  the deviations from the DESI data and 
finally we estimate the $\chi_{\rm min}^2$ from the given covariance matrix 
which comprises the   statistical and systematic errors. The latest version of 
DESI BAO measurement has been enlisted in DR2 
dataset~\cite{DESI:2025zgx}.\footnote{\url{
https://github.com/CobayaSampler/bao_data/tree/master/desi_bao_dr2}} In contrast 
to DR1, where the data of ${D_{\rm V}}/{r_{\rm d}}$ at redshift 1.491 was used, 
there are the measured values of ${D_{\rm M}}/{r_{\rm d}}$ and ${D_{\rm 
H}}/{r_{\rm d}}$ at redshift 1.484. Hence, in total 13 data points of 
distance-to-sound-horizon ratio from 7 red-shift bins are present in DESI DR2. 
These data sets also give the latest covariance among the observed quantities.
The likelihood function is, $\mathcal{L} = e^{-\chi^2/2}$. And $\chi^2$ is calculated as, $\chi^2 = \Delta D_{\rm i}^{\rm T}\left(\mathcal{C}_{\rm DESI}^{-1}\right)_{\rm ij}\Delta D_{\rm j}$, where $\Delta D_{\rm i} = \left(D_{\rm i}/r_{\rm d}\right)_{\rm data} - \left(D_{\rm i}/r_{\rm d}\right)_{\rm model}$, and `$i$' denotes $M$ (corresponds to the measurements of $D_M/r_d$), $V$ (corresponds to the measurements of $D_V/r_d$) and $H$ (corresponds to the measurements of $D_H/r_d$) according to the order of the 7 redshift binning DESI data. The covariance matrix $\mathcal{C}_{\rm DESI}$ is used in the same order. Then we have calculated $\chi_{\rm DESI}^2$.

\end{enumerate}

In order to constrain the model parameters, we perform the Markov Chain Monte 
Carlo (MCMC) analysis using the publicly available python package \texttt{emcee 
package}~\cite{Foreman-Mackey:2012any}.\footnote{\url{
https://emcee.readthedocs.io}} 
We consider a sufficient number of iterations until  the auto-correlation time 
gets saturated and the MCMC chains get 
converged~\cite{Goodman:2010dyf,10.1093/oso/9780198522669.003.0038}. For the 
analysis and visualization of the final MCMC chains, we use the  
\textbf{\texttt{getdist}} 
package~\cite{Lewis:2019xzd}.\footnote{\url{https://github.com/cmbant/getdist}} 
Now, since baryons and radiation are very insignificant at present time, 
 for both Model I and Model II  we   fix  $\Omega_{\rm b,0} \sim 
0.0493$ and $\Omega_{\rm r,0} \sim 9.24\times10^{-5}$ from Planck 
2018~\cite{Planck:2018vyg} and Particle Data Group 
(PDG)~\cite{ParticleDataGroup:2024cfk}, and as a result  one can derive 
$\Omega_{\rm dm,0}$ using the relation $\Omega_{\rm dm,0} = 1 - \Omega_{\rm 
b,0}-\Omega_{\rm r,0} = 1 - 0.0493 - 9.24\times10^{-5} \approx 0.9506$. Thus, for 
Model I, we have two free parameters, namely  $\alpha$ and $H_0$, while for 
Model II we have three free parameters, namely $\alpha$, $l$ and $H_0$. 
The priors on the parameters are as follows:  $H_0 \thicksim \mathcal{U}(60, 
80)$, $\alpha \thicksim \mathcal{U}(0,1)$ and $l \thicksim \mathcal{U}(-1,1.2)$ 
($l \neq 0$). We use a flat prior range for $\mathcal{M}$ and let it pass 
through the MCMC samplings, namely  $\mathcal{M} 
\thicksim \mathcal{U}(22.5,24.5)$.

\section{Results}
\label{sec-results}

In this section we summarize the 68\% confidence level (CL) constraints on the 
model parameters of Model I and Model II, using the above observational 
datasets. Moreover, we test the performance of the present models using 
various model comparison statistics. For this purpose, we   consider the 
standard $\Lambda$CDM scenario as the base model.  In the following subsections 
we present the main results.

\subsection{Observational Constraints}

\begin{figure}
    \centering
    \includegraphics[width=0.94\linewidth]{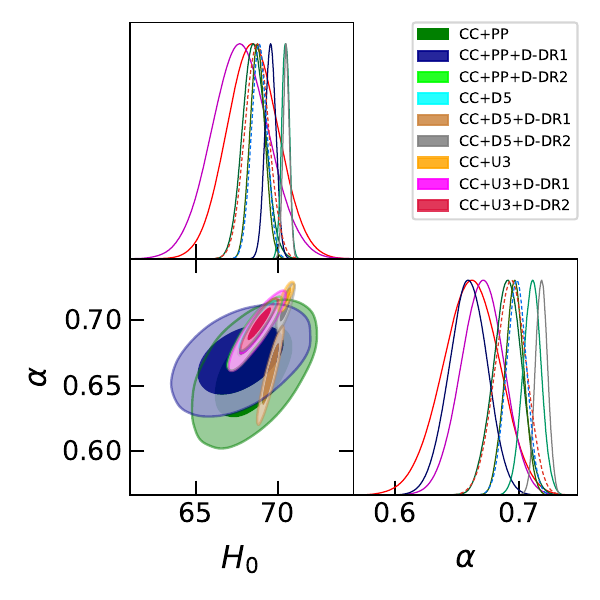}
    \caption{{\it{The posterior distribution and the two-dimensional joint 
likelihood 
contours of Model I for all the data 
combinations.}}}
    \label{fig-M1-Trian}
\end{figure}

\begin{figure}
    \centering
    \includegraphics[width=0.94\linewidth]{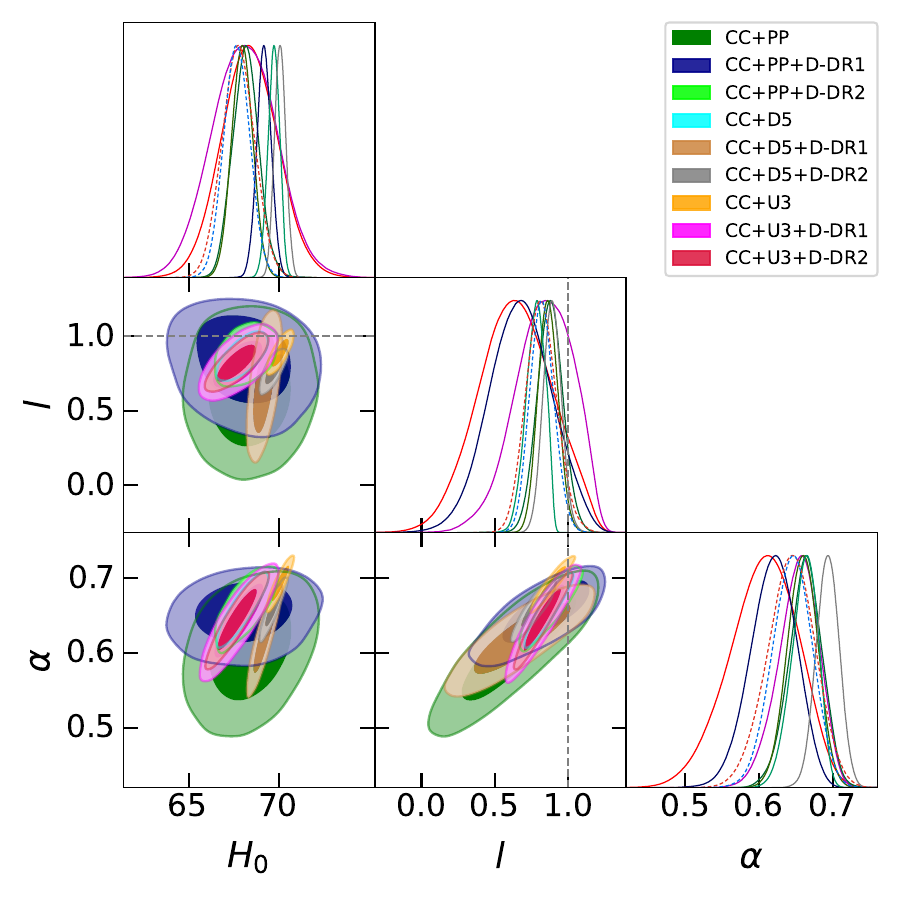}
    \caption{{\it{The posterior distribution and the two-dimensional joint 
likelihood 
contours  of Model II for all the data 
combinations.}}}
    \label{fig-M2-Trian}
\end{figure}

In Tables~\ref{Tab-data-M1} and~\ref{Tab-data-M2} we respectively present  the 
constraints on the model parameters of Model I and II.  Additionally, in 
Figs.~\ref{fig-M1-Trian} and~\ref{fig-M2-Trian} we show the one-dimensional 
marginalized posterior distributions and the two-dimensional joint likelihood 
contours corresponding to Model I and Model II, respectively. Furthermore, in 
Figs.~\ref{fig-dat-M1} and~\ref{fig-dat-M2} we specifically demonstrate the 
effects of DESI DR1 and DESI DR2 on the combined dataset CC+SNIa. Recall that 
for both the models, $\alpha \neq 0$ indicates the evidence of matter creation, 
while for Model II $l \neq 1$  indicates that the resulting scenario deviates 
from   $\Lambda$CDM cosmology since $l =1$ leads to the $\Lambda$CDM-like 
evolution (see Eq.~\ref{m1_data}). $\Omega_{\rm dm, 0 }^{\rm eff}$ reads from Eq.~\eqref{m1_data}, $\Omega_{\rm dm, 0 }^{\rm eff} = (\Omega_{\rm dm,0} - \alpha)$, 
and the effective constant energy density parameter, $\Omega_{\Lambda}^{\rm eff} = \alpha$. $\Omega_{\rm dm, 0 }^{\rm eff}$ reads from Eq.~\eqref{m2_data}, $\Omega_{\rm dm, 0 }^{\rm eff}\simeq (\Omega_{\rm dm,0}^l -\alpha)^{1/l} $,
and the effective constant energy density parameter, $\Omega_{\Lambda}^{\rm eff} \simeq \alpha^{1/l}$.
Let us analyze the various cases in 
detail.

\begin{enumerate}
    \item \textbf{CC+PP}: 
    For Model I,  for this dataset we find a clear evidence for non-null $\alpha 
$ yielding $\alpha = 0.670 \pm 0.018$ at 68\% CL. The Hubble 
constant assumes similar value as noticed in the $\Lambda$CDM based Planck's 
case~\cite{Planck:2018vyg} but with slightly higher error bars. 

    For Model II  evidence of matter creation is also found at several standard 
deviations similar to what we noticed in Model I. However, in this case we find 
that $l$ allows $1$ within 68\% CL, which   implies that the deviation from 
  $\Lambda$CDM cosmology is statistically weak. Moreover, the Hubble 
constant assumes  slightly higher values ($H_0 = 68.03^{+0.57}_{-0.67}$ 
km/s/Mpc at 68\% CL) compared to Planck  (within $\Lambda$CDM 
paradigm)~\cite{Planck:2018vyg}, nevertheless  not significantly.  
    Overall, for both the models, CC+PP dataset favors   matter creation.

\begin{figure*}[t]
    \centering
    
\subfigure[]{\includegraphics[width=0.32\textwidth]{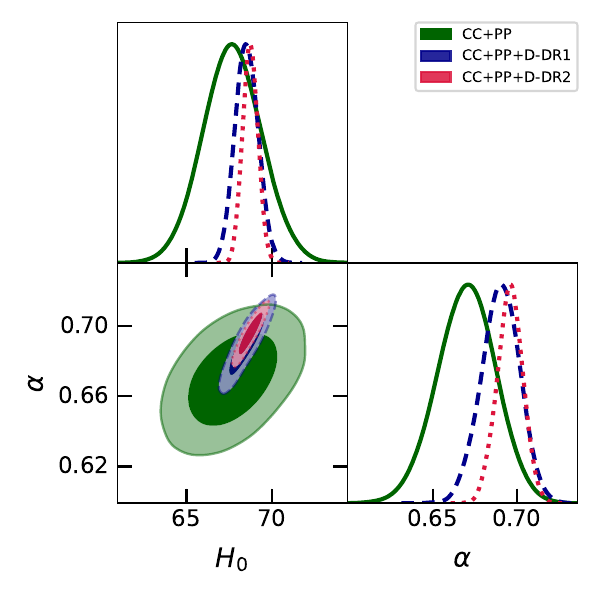}}\label
{fig-M1-PP}    
\subfigure[]{\includegraphics[width=0.32\textwidth]{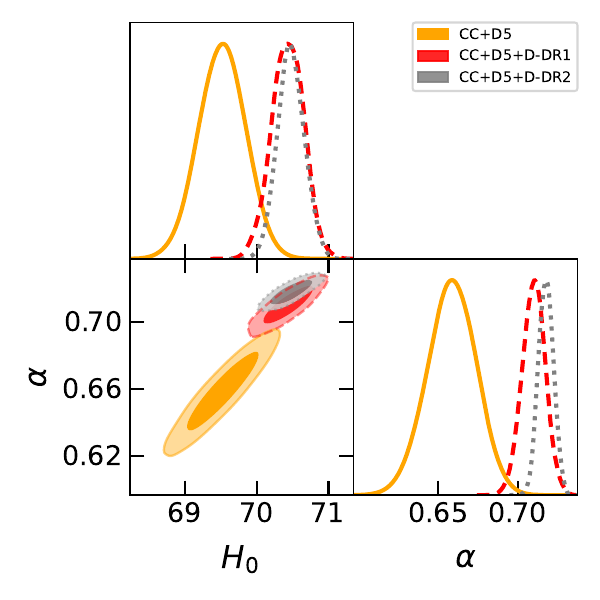}}\label
{fig-M1-D5}    
\subfigure[]{\includegraphics[width=0.32\textwidth]{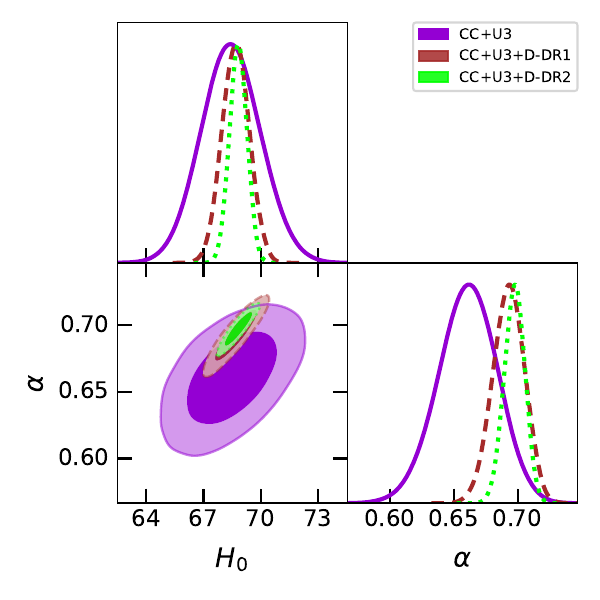}}\label
{fig-M1-U3}
    \caption{Posterior distributions and   two-dimensional joint likelihood 
contours  for different supernovae data with the DESI DR1 and DR2 effects for 
Model I. }
    \label{fig-dat-M1}
\end{figure*}    

\begin{figure*}[t]
    \centering
    
\subfigure[]{\includegraphics[width=0.32\textwidth]{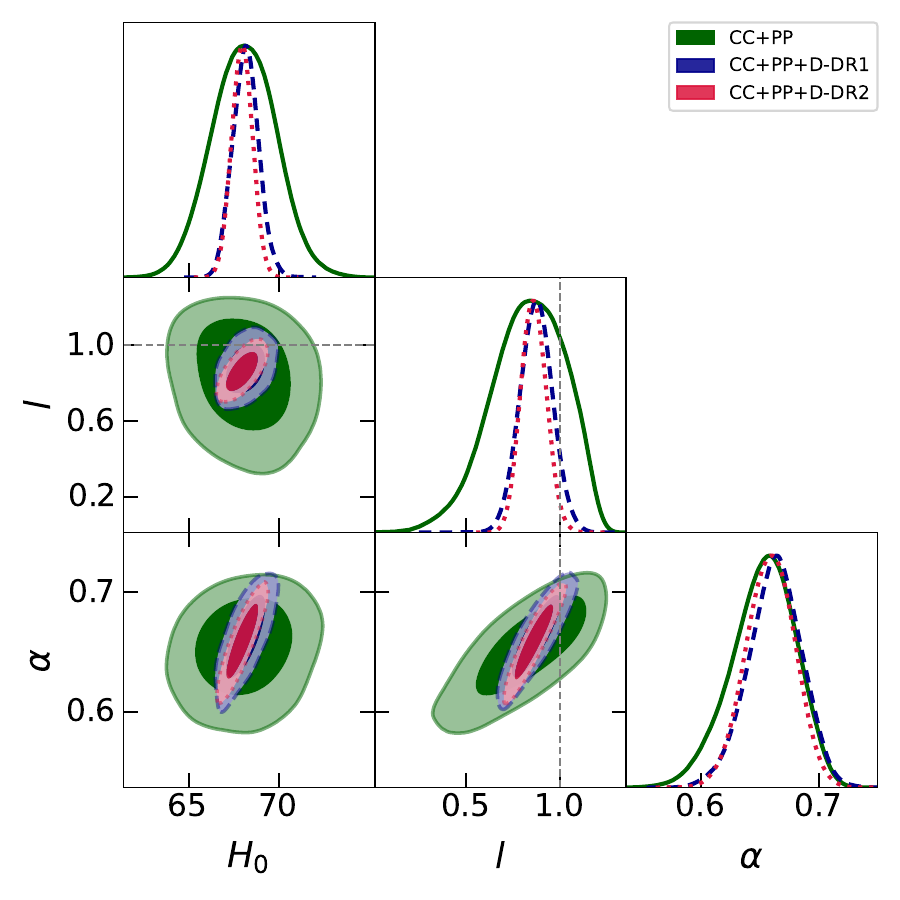}}\label
{fig-M2-PP}    
\subfigure[]{\includegraphics[width=0.32\textwidth]{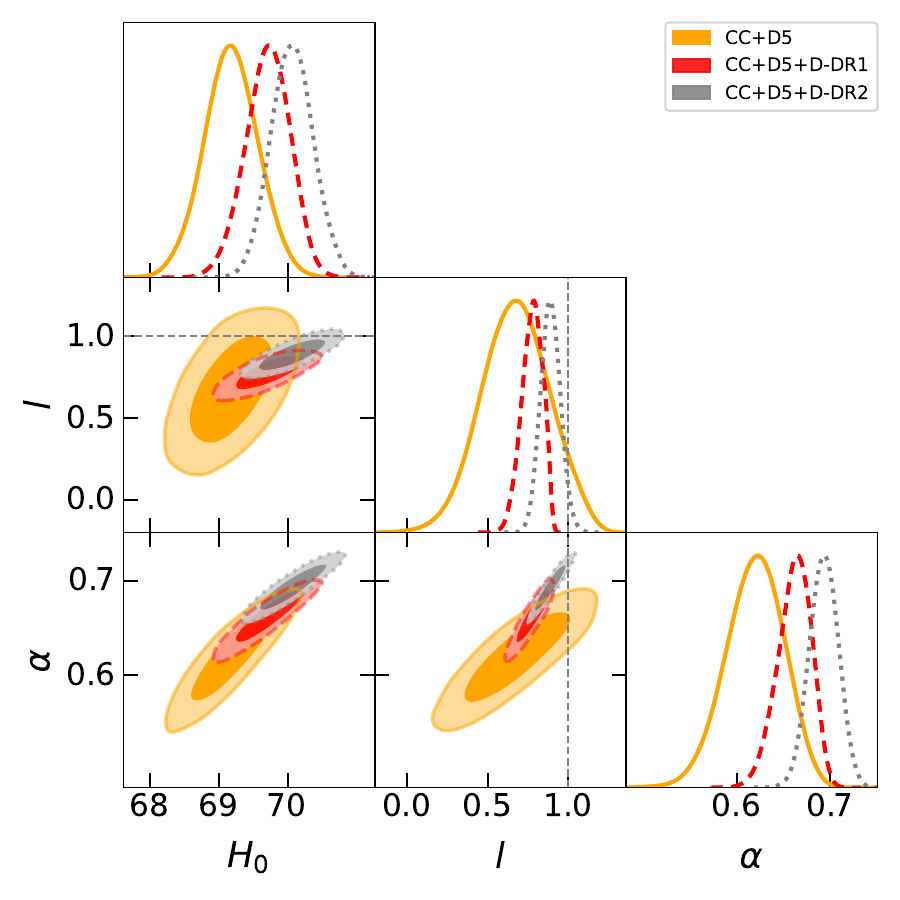}}\label
{fig-M2-D5}    
\subfigure[]{\includegraphics[width=0.32\textwidth]{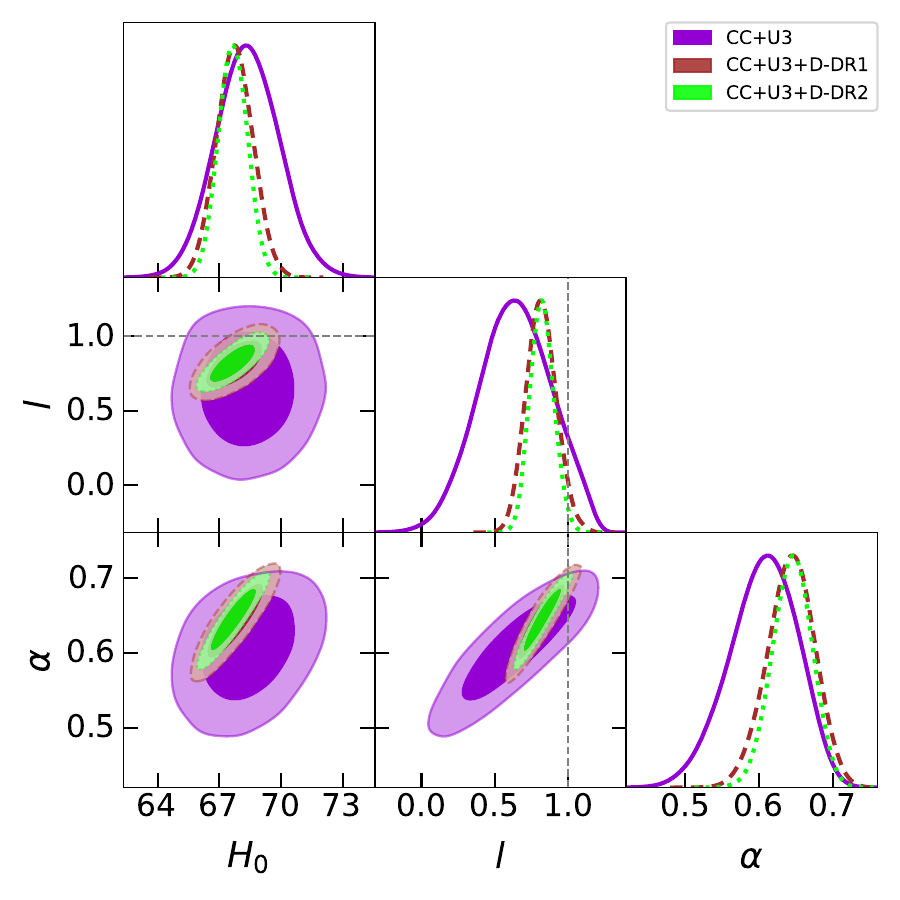}}\label
{fig-M2-U3}
    \caption{Posterior distributions and   two-dimensional joint likelihood 
contours  for different supernovae data with the DESI DR1 and DR2 effects for 
Model II. }
    \label{fig-dat-M2}
\end{figure*}

    \item \textbf{CC+PP+DESI-DR1}:
    The inclusion of DESI-DR1 with CC+PP affects the constraints on both   
models. For Model I   we notice   changes in the mean values of the 
parameters as well as the reduction in the uncertainties. The maximum reduction 
in uncertainties is found in $H_0$.     However, we find the evidence of matter 
creation in terms of non-null $\alpha$. The left graph of Fig. 
\ref{fig-dat-M1} clearly shows the impact of DESI-DR1 when combined with CC+PP. 
 
  For Model II (see the left graph of Fig. \ref{fig-dat-M2}), one can similarly 
observe the effects of DESI-DR1. Specifically, we observe that the constraint 
on $H_0$ does not change but its error bars are reduced by a factor of 2.5 which 
is significant. For the remaining two parameters we find mild shift in their 
mean values, however their uncertainties are reduced. Furthermore, we still 
find an evidence of matter creation at several standard deviations, and $l$ 
also deviates from $1$ at slightly more than 68\% CL. Hence, DESI-DR1 increases 
the deviation from   $\Lambda$CDM cosmology, nevertheless in this case such 
deviation is not statistically robust.

\item \textbf{CC+PP+DESI-DR2}:
When DESI-DR2 is combined with CC+PP (alternatively, when DESI-DR1 is replaced 
by DESI-DR2 in the combined dataset CC+PP+DESI-DR1), we can similarly observe 
that the constraining power of the combined dataset increases because of 
DESI-DR2. Comparing the constraints from CC+PP and CC+PP+DESI-DR2 obtained in 
Model I, we see that the mean values of $H_0$ and $\alpha$ slightly change but 
the significant changes appear in terms of the reduction of their uncertainties. 
In particular, the uncertainty in $H_0$ is reduced by a factor of $3.8$, and in 
case of $\alpha$  its uncertainty is reduced by a  factor of $2.3$. One can 
also notice that in this case DESI-DR2 shows better constraining power than 
DESI-DR1. 
Overall, the evidence of matter creation is strengthened in this case, and this 
can be seen in the left graph of Fig. \ref{fig-dat-M1}.

For Model II   our observation is almost similar to Model I. This implies that  
the constraints on all the parameters slightly change, but significant changes 
appear in the reduction of their uncertainties. We find the evidence of matter 
creation at more than 68\% CL, but interestingly  we notice that $l$ now 
deviates from $1$ at  95\% CL. From this we deduce that the deviation from   
$\Lambda$CDM cosmology is    strengthened for this combined dataset (see the 
left graph  of Fig. \ref{fig-dat-M2} for the effect of  DESI-DR2  dataset when 
combined with CC+PP).

\item \textbf{CC+D5}: 
For both Model I and Model II we find evidence of matter creation quantified by 
non-zero $\alpha$  at several standard deviations. As we see, the estimated 
values of the Hubble constant for both models  are slightly higher than the 
$\Lambda$CDM based Planck's estimation~\cite{Planck:2018vyg}.
Additionally, for 
Model II, which has one extra free parameter, $l$ is also found to be deviating 
from $1$ at more than 95\% CL, which implies that  the deviation from   
$\Lambda$CDM cosmology is favored.

\item \textbf{CC+D5+DESI-DR1}: 
The inclusion of DESI-DR1 with CC+D5 affects the parameter space of both   
models.  For Model I, the mean values of $H_0$ and $\alpha$ slightly shift with 
some reduction in their uncertainties  (see the middle graph of 
Fig. \ref{fig-dat-M1}), and  in particular the uncertainty of $\alpha$ is 
reduced by a factor of $2$. However,  evidence 
of matter creation persists and rather strengthens, due to the reduction of 
its uncertainty.

For Model II we observe mild shifts in the mean values of the model 
parameters (see the middle graph of Fig. \ref{fig-dat-M2}), while we notice 
maximum improvements  through the reduction in the uncertainties of 
the involved parameters. 
Overall, we notice evidence of matter creation,  
and the parameter $l$ deviates from $1$ at several standard deviations,
suggesting the possible deviation from the $\Lambda$CDM cosmology. 

We would like to note that the estimated values of $H_0$ in both the models  are slightly higher ($H_0 \sim 69.7 - 70.4$ km/s/Mpc) than the Planck $\Lambda$CDM values~\cite{Planck:2018vyg}. In fact, comparing the values of $H_0$ obtained in other datasets (see Tables~\ref{Table-M1} and \ref{Table-M2}), it is evident that $H_0$ takes slightly higher values when  D5 dataset is present in the combined dataset (for CC+D5, $H_0 \sim 69.2 - 69.5$ km/s/Mpc). This tells that D5 dataset can play an effective role in alleviating the Hubble constant tension, compared to the other SNIa datasets in this context. While the shift toward higher $H_0$ is not fully driven by the matter creation scenario, it is also a  dataset-driven effect, reflecting the distinct redshift coverage and covariance properties of the D5 dataset, which may alter the relative sensitivity to the late-time expansion history and consequently shift the preferred value of $H_0$
 compared to the other two SNIa datasets, namely PP and Union3. A full analysis using the cosmic microwave background radiation from Planck and other astronomical surveys can offer more complete information in this direction.

\item \textbf{CC+D5+DESI-DR2}: 
The constraints for Model I between CC+D5+DESI-DR1 and CC+D5+DESI-DR2 are almost 
identical except some improvements in the uncertainties (see the middle graph of 
Fig. \ref{fig-dat-M1}). The Hubble constant takes slightly higher value
($H_0 \sim 70$ km/s/Mpc), and thus helps to reduce the tension  a bit. The uncertainty on $H_0$ for CC+D5+DESI-DR2 is reduced 
by a factor of $1.7$ and the uncertainty on $\alpha$ is reduced by a factor of 
$\sim 3$. This shows that for this model, DESI-DR2 adds much constraining power 
compared to DESI-DR1.   
Overall, we find that the evidence of matter creation  
is strengthened compared to CC+D5 and CC+D5+DESI-DR1.  

For Model II  we first note that the inclusion of DESI-DR2 with CC+D5, affects 
the constraints. The uncertainty in $\alpha$ is reduced compared to CC+D5 and 
the evidence of matter creation is similarly observed as in other cases. The Hubble constant is mildly uplifted ($H_0 \sim 70.1$ km/s/Mpc) compared to CC+D5+D-DR1 yielding $H_0 \sim 69.7$ km/s/Mpc.
However, in this case we find that the mean value of $l$ increases and shifts 
toward $1$, although $l \neq 1$ is found at slightly greater than 68\% CL but 
less than 95\% CL. This can be explained though the correlations between $\alpha 
- H_0 - l$ as shown in the middle graph of Fig. \ref{fig-dat-M2}. As $\alpha$, 
and $l$ are positively correlated with $H_0$,  an  increase in $\alpha$ 
increases the value of $H_0$ which results to a higher value of $l$.

\item \textbf{CC+U3}:
According to the results, irrespective of the models, we find  evidence of 
matter creation at several standard deviations. The Hubble constant for 
both Model I and Model II is estimated to be slightly higher ($H_0 \sim 68.5$ 
km/s/Mpc) compared to the Planck 2018 (within the $\Lambda$CDM 
paradigm)~\cite{Planck:2018vyg}. Additionally, for Model II, although the mean 
value of $l$ differs from $1$, the value $ l= 1$ is allowed within 68\% CL.

\item \textbf{CC+U3+DESI-DR1}: 
In this case we find that for both   
models  the constraints on the model parameters are  improved, due to  the 
reduction in  the uncertainties (see the right graph of Fig. \ref{fig-dat-M1}). 
For Model I we only notice a mild shift in $\alpha$, however  its    
uncertainty is reduced by a factor of $\sim 1.9$, while  the uncertainty in 
$H_0$ is reduced by a   factor of $2.2$. We find evidence of matter 
creation     at several standard deviations.  
   
For Model II we find an improvement in the uncertainty of $H_0$ by a factor 
of $\sim 1.8$. The mean values of $l$ and $\alpha$ increase compared to CC+U3 
but their uncertainties are reduced compared to CC+U3 (see the right graph of 
Fig. \ref{fig-dat-M2}). Overall, we find   evidence of matter creation,  and we 
notice that  $l \neq 1$  at slightly more than 68\% CL. Thus, in this case the
deviation from   $\Lambda$CDM cosmology is mildly suggested.

 \item \textbf{CC+U3+DESI-DR2}:
    The inclusion of DESI-DR2 with CC+U3 similarly affects the parameter space 
by reducing the uncertainties of $H_0$ and $\alpha$. In comparison to DESI-DR1, 
DESI-DR2 has slightly better constraining power, which is reflected through the 
reduction in the uncertainties of the parameters. In particular,  for Model I, 
the mean value of $\alpha$ is slightly changed compared to CC+U3, but it 
remains almost identical with CC+U3+DESI-DR1. The uncertainty on $\alpha$ is 
reduced by a factor of $2.8$ and for $H_0$ its uncertainty is reduced by a 
factor of $2.2$. 

 For Model II, the  DESI-DR2  inclusion in   CC+U3, 
  offers stringent constraints by reducing 
the uncertainties of the parameters. However, in this case, the mean values of 
the parameters obtained for CC+U3+DESI-DR1 and CC+U3+DESI-DR2 remain almost 
identical (see the last graph of Fig. \ref{fig-dat-M2}), but again we
observe  that the constraining power of DESI-DR2 is higher than DESI-DR1.  
Overall, we find evidence of matter creation   at several standard 
deviations, which suggests a deviation from   $\Lambda$CDM 
cosmology at slightly more than 95\% CL.

\end{enumerate}

\subsection{Model comparison}

In order to assess the acceptability of  the models, in this subsection we 
perform a model comparison analysis considering the well known statistics. As a 
reference model we consider the $\Lambda$CDM 
  scenario. 
We apply the  Akaike Information Criterion (AIC) defined as
$\mathrm{AIC} = \chi_{\mathrm{min}}^2 + 2k$  \cite{1100705}, and the Bayesian 
Information Criterion (BIC) defined as  $\mathrm{BIC} = \chi_{\mathrm{min}}^2 + 
k \ln(N)$ ~\cite{10.1214/aos/1176344136, 
Trotta:2008qt}, where $k$ refers to the total number of free parameters and $N$ 
is the total number of data points.  For comparison between the models we 
calculate  $\Delta \chi_{\rm min}^2$, $\Delta \rm{AIC}$ and $\Delta \rm BIC$ 
with respect to the standard $\Lambda$CDM model through  
  $\Delta \chi_{\rm min}^2 = \chi_{\rm min}^2|_{\Lambda \rm{CDM}} - 
\chi_{\rm min}^2|_{\rm model} $, $\Delta \rm{AIC} = \rm{AIC}_{\Lambda \rm CDM} 
- \rm{AIC}_{\rm model} $ and $\Delta \rm{BIC} = \rm{BIC}_{\Lambda CDM} - 
\rm{BIC}_{model} $.  

The strength of evidence can be described as follows 
\cite{burnham2002model}: 
$0<\Delta \rm{AIC}<2$ suggests  statistical equivalence  for the test 
model against the reference model;  $4<\Delta \rm{AIC}<7$ implies a 
considerable strength for the test model, while $\Delta \rm{AIC}>10$ suggests 
very strong evidence for the test model. Concerning BIC, one can describe the 
strength of evidence as follows~\cite{Kass:1995loi}: $0<\Delta \rm{BIC} < 2$ 
suggests weak or equal evidence of the test model with respect to the reference 
model; $2<\Delta \rm{BIC} < 6$ suggests moderate evidence for test model; 
$6<\Delta \rm{BIC} < 10$ and $\Delta \rm{BIC} > 10$ suggest strong to very 
strong evidences for the test model. 

Comparing Model I and $\Lambda$CDM scenario  we see that the number of 
parameters are equal for each model, thus  the preference of Model I with 
respect to $\Lambda$CDM is achieved through $ \Delta \chi_{\rm min}^2> 0$.  
 Since the number of parameters and used data points are the same for a 
particular combined datasets, $\Delta \rm{AIC}$ and $\Delta \rm{BIC}$  are both 
identical to $ \Delta \chi_{\rm min}^2$. Therefore, $ \Delta \chi_{\rm min}^2> 
0$ implies that both $\Delta \rm{AIC}$ and $\Delta \rm{BIC}$  are 
positive. Hence Model I is favored over the reference 
model under this criteria. 

On the other hand, while comparing Model II  with $\Lambda$CDM scenario, there 
is one extra parameter `$l$' (apart from the case  $l=1$  where the results are 
analogous to $\Lambda $CDM (see Eq.~\ref{m1-rho-dm-sol}). Thus,  $\Lambda $CDM 
scenario is nested within Model II. Hence, we can apply the Likelihood Ratio 
Test (LRT) while comparing the $\Delta \chi^2_{\rm min}$ between our model and 
$\Lambda $CDM. LRT states that, $\Delta \chi_{\rm min}^2$ has to be greater than 
3.84146 to favor the test model at 95\% CL~\cite{Montanari:2024dxh, 
Corstanje:2021kik}. On the other hand, since $\Delta \mathrm{AIC} = \Delta 
\chi^2_{\min} + 2\,\Delta k$ and $\Delta \mathrm{BIC} = \Delta \chi^2_{\min} + 
\Delta k \ln(N)$, and in our case $\Delta k = -1$ (i.e., the test model has one 
additional parameter compared to the reference model), the expressions reduce 
to:
\[
\Delta \mathrm{AIC} = \Delta \chi^2_{\min} - 2, \qquad \Delta \mathrm{BIC} = \Delta \chi^2_{\min} - \ln(N).
\]
In order for the information criteria to favor the test model  the minimum 
requirement is  
\[
\Delta \chi^2_{\min} > 2 \quad \text{and} \quad \Delta \chi^2_{\min} > \ln(N),
\]
respectively.

In Table~\ref{tab-bic-penalties}  we provide the values of $\ln(N)$ for various 
combinations of datasets. This illustrates that, for data combinations with a 
large number of points, the required improvement in $\chi^2_{\min}$ must be 
correspondingly larger to overcome the BIC penalty associated with the 
complexity and sample size of the model. Therefore, a significant $\Delta 
\chi^2_{\min}$ is necessary for the test model to be favored under the BIC 
criterion in such cases.

\begin{table}[t]
\centering
\begin{tabular}{ccc}
\hline
Datasets & $N$ &$\ln(N)$\\
\hline
CC + PP &1624 &$7.39$ \\
CC + PP + D-DR1&1636 &$7.40$\\
CC+PP+D-DR2&1637 &$ 7.40$ \\
CC+D5&1788 &$7.48$ \\
CC+D5+D-DR1&1800 &$7.49$ \\
CC+D5+D-DR2&1800 &$7.49$  \\
CC+U3  & 56 &$4.02$\\
CC+U3+D-DR1& 68 &$4.22$ \\
CC+U3+D-DR2&69 &$4.23$ \\
\hline
\end{tabular}
\caption{Values of $\ln(N)$ for each data combination, representing the $\rm 
BIC$ penalty per additional parameter (see text).}
\label{tab-bic-penalties}
\end{table}

In
Fig. \ref{fig-aic-bic} we graphically display the statistical behavior of the 
models considering the same statistical quantities. 
 Additionally, in  Tables~\ref{tab-m1-lcdm-chi2} and  \ref{tab-m2-lcdm-chi2} 
we summarize the 
values of $ \Delta \chi_{\rm min}^2$, $\Delta \rm{AIC}$ and $\Delta \rm{BIC}$  
for both   models, considering all the combined datasets. 

From Table~\ref{tab-m1-lcdm-chi2}  and 
the upper graph of Fig. \ref{fig-aic-bic}
one can see that for all   data combinations, Model I and $\Lambda $CDM are in 
same preference footing for all statistical quantities (e.g. $\Delta 
\chi^2_{\rm 
min}$, $\Delta \rm{AIC}$ and $\Delta \rm{BIC}$)
except for the combinations  CC + PP + DESI-DR2 and CC + DESY5 + DESI-DR1, 
which have a slight edge to the reference model $\Lambda$CDM, however  this is 
not significant. Overall, Model I is found to be very close to   $\Lambda$CDM 
scenario, i.e.  they are practically indistinguishable. 

From Table~\ref{tab-m2-lcdm-chi2}  and the lower graph of Fig. 
\ref{fig-aic-bic}, 
we can notice that for the  dataset  combinations CC+PP+D-DR2, CC+D5+D-DR1 and 
CC+U3+D-DR2,  Model II is preferable over $\Lambda $CDM by means of LRT 
($\Delta \chi^2_{\rm min} >3.84 \implies 95\%$ CL for test model) and $\Delta 
\rm{AIC}$. The values of $\Delta \rm{AIC}$ for all dataset combinations provide 
a comparable result between $\Lambda $CDM and Model II, except for CC+PP and 
CC+U3. By means of $\Delta \rm{BIC}$, we have only CC+U3+D-DR2, which has a 
slight edge for the test model, but for the remaining dataset combinations,  
$\Lambda$CDM is indeed preferable. In spite of these, we can notice an 
increasing trend of $\Delta \rm BIC$ for all of the CC+SNIa combinations.

\section{Summary and Conclusions}\label{sec-summary}

Cosmological  matter creation scenarios are considered to be an alternative 
approach to both dark energy and modified gravity theories. In such scenarios, 
it is commonly assumed that the DM sector is responsible for creating matter 
particles, and as a result  DM does not follow the standard conservation 
law. Due to the violation of the standard  conservation law, a creation 
pressure appears, which depends on the creation rate of the 
matter particles, and thus the creation rate can affect the expansion history of 
the universe. Hence under  a proper choice of   creation rate one can obtain 
the present accelerating phase as well as the past thermal history of the 
Universe.  In this work we have studied two matter creation scenarios,
namely Model I, characterized by $\Gamma = 3 \alpha H 
(\rho_{\rm c0}/\rho_{\rm dm})$, and its extension Model II, characterized by  
$\Gamma = 3 \alpha H (\rho_{\rm c0}/\rho_{\rm dm})^l$.

Firstly we applied the powerful method of dynamical analysis, which allows to 
extract information on the global behavior of the system, independently of the 
initial conditions. As we showed,  Model I   
exhibits the sequence of radiation, matter 
and late-time acceleration epochs, and that it can moreover mimic a   
$\Lambda$CDM-like behavior.
In the case of Model II, since  it involves the additional free parameter $l$, 
we considered four $l$  ranges  separately. We obtained  a transient 
phantom-like evolution, a $\Lambda$CDM-like 
evolution, and a phantom-like evolution.  However, since the trajectories which 
present transient phantom-like behavior and phantom-like behavior are 
suffering from early acceleration just from the beginning of matter 
domination,   they are not physically interesting, leaving the 
$\Lambda$CDM-like behavior the only interesting solution.
 
In order to obtain a full picture for the behavior and the efficiency of the 
matter creation models,   we performed a detailed observational confrontation 
using   a series of latest observational datasets including  CC, SNIa 
(Pantheon+, DESY5 and Union3) and DESI BAO (DR1 and DR2). 
As showed, we found an evidence of matter creation in 
both   models at many standard deviations. In the case of Model II, 
we found that in most     datasets 
$l$ also deviates from the value $1$, hence a deviation from the standard 
$\Lambda$CDM cosmology is favored. 
This is a significant  result since  the 
recent DESI-DR2 results also indicate an evidence of dynamical DE and hence a 
deviation from  standard $\Lambda$CDM cosmology. We have also noticed that in the presence of DESY5 in the combined datasets, the estimated values of $H_0$ are slightly higher compared to the other two SNIa datasets. While this result may have potential implications in supporting matter creation models as a possible mechanism to alleviate the Hubble constant tension, however, as the present article is restricted to the analyses at the background level, a more robust conclusion would require the inclusion of additional observational probes such as the cosmic microwave background data.

Additionally, we performed a comparison analysis with $\Lambda$CDM paradigm,  
considering the AIC and BIC information criteria. As we showed,
 Model I and $\Lambda$CDM scenario are almost indistinguishable, apart from  the 
cases CC+PP+D-DR2 and CC+D5+D-DR1 according to which both AIC and BIC favor 
slightly  $\Lambda$CDM. 
On the other hand,   the comparison between Model II and $\Lambda$CDM 
scenario shows a mixed picture, namely for most datasets  $\Lambda$CDM scenario 
is favored, however,  DESI data with CC+SNIa combination favor matter 
creation Model II over   $\Lambda$CDM paradigm.

In summary, we showed that matter creation can be an alternative scenario to 
both dark energy and modified gravity, which can potentially be favored 
comparing to $\Lambda$CDM paradigm. Definitely, one should extend the analysis 
beyond the background 
level, and analyze the perturbations too, confronting them with  the full 
cosmic microwave background data.  As the matter creation mechanism introduces the creation pressure through the rate of matter creation $\Gamma$, therefore, it leads to an effective DM EoS which modifies the expansion history of the universe at both the background and perturbative level. As a result, it is expected that the growth of structure of DM should alter compared to the standard growth of structure of DM in absence of the matter creation. We anticipate that 
the inclusion of cosmic microwave background data and other astronomical probes could offer stringent 
constraints on the models, nevertheless since such a detailed analysis lies 
beyond the scope of this first work on the subject, it is left for a future 
project.
 \section*{Acknowledgements}
  We thank the referees for several important comments which helped us to improve the quality of the manuscript. SB acknowledges Indian Statistical Institute, Kolkata for the financial grant 
(Office Order No. DS/2024-25/0616 Dated 23 August 2024). Also, SB thanks the 
Department of Mathematics, Presidency University, Kolkata where a substantial 
part of the work has been carried out. We thank Prof. Subir Ghosh for his 
valuable remarks on the article.
SH acknowledges the financial support from the University Grants Commission 
(UGC), Govt. of India (NTA Ref. No: 201610019097). JdH is supported by the 
Spanish grants PID2021-123903NB-I00 and RED2022-134784-T
funded by MCIN/AEI/10.13039/501100011033 and by ERDF ``A way of making 
Europe''. 
SP acknowledges the partial support from the Department of Science and 
Technology (DST), Govt. of India under the Scheme   ``Fund for Improvement of 
S\&T Infrastructure (FIST)'' (File No. SR/FST/MS-I/2019/41). ENS acknowledges 
the contribution of the LISA 
CosWG, and of   COST 
Actions  CA18108  "Quantum Gravity Phenomenology in the multi-messenger 
approach''  and  CA21136" Addressing observational tensions in cosmology with 
systematics and fundamental physics (CosmoVerse)''.  We acknowledge the 
computational cluster resources at the ICT Department of Presidency University, 
Kolkata.

\appendix

\section{Model I}\label{appen-A}

The continuity equation for the dark matter energy density $\rho_{\rm dm}$ is (cf. Eq.~(\ref{eq20}))
\begin{equation}
\dot{\rho}_{\rm dm} + 3H\rho_{\rm dm} = 3\alpha H \left(\frac{\rho_{\rm c,0}}{\rho_{\rm dm}}\right)\rho_{\rm dm}.
\label{A1}
\end{equation}
This simplifies to
\begin{equation}
\dot{\rho}_{\rm dm} + 3H\rho_{\rm dm} = 3\alpha H \rho_{\rm c,0},
\end{equation}
or equivalently,
\begin{equation}
\dot{\rho}_{\rm dm} + 3H(\rho_{\rm dm} - \alpha \rho_{\rm c,0}) = 0.
\end{equation}

Using $\dot{\rho}_{\rm dm} = \dot{a}\,\frac{d\rho_{\rm dm}}{da}$, we rewrite the above equation as
\begin{equation}
\frac{d\rho_{\rm dm}}{\rho_{\rm dm} - \alpha \rho_{\rm c,0}} = -3\,\frac{da}{a}.
\end{equation}
Integrating both sides, we obtain
\begin{equation}
\ln\left|\rho_{\rm dm} - \alpha \rho_{\rm c,0}\right| = -3\ln a + C,
\label{A2}
\end{equation}
where $C$ is an integration constant. Imposing the present-day condition $a = 1$ and $\rho_{\rm dm} = \rho_{\rm dm,0}$, we find $C = \ln|\rho_{\rm dm,0} - \alpha \rho_{\rm c,0}|$. Therefore,
\begin{equation}
\rho_{\rm dm} = \alpha \rho_{\rm c,0} + (\rho_{\rm dm,0} - \alpha \rho_{\rm c,0})a^{-3}.
\label{A3}
\end{equation}

In terms of the redshift $z = \frac{1-a}{a}$, this becomes
\begin{equation}
\frac{\rho_{\rm dm}}{\rho_{\rm c,0}} = \alpha + \left(\frac{\rho_{\rm dm,0}}{\rho_{\rm c,0}} - \alpha\right)(1+z)^3,
\end{equation}
or
\begin{equation}
\frac{\rho_{\rm dm}}{\rho_{\rm c,0}} = \alpha + (\Omega_{\rm dm,0} - \alpha)(1+z)^3,
\label{A5}
\end{equation}
where $\Omega_{\rm dm,0} \equiv \frac{\rho_{\rm dm,0}}{\rho_{\rm c,0}}$.

From Eq.~(\ref{friedmann-1A}), the Friedmann equation can be written as
\begin{equation}
\left(\frac{H}{H_0}\right)^2 = \frac{\rho_{\rm dm}}{\rho_{\rm c,0}} + \frac{\rho_{\rm b}}{\rho_{\rm c,0}} + \frac{\rho_{\rm r}}{\rho_{\rm c,0}}.
\label{A6}
\end{equation}
Using $\rho_{\rm b} = \rho_{\rm b,0}(1+z)^3$, $\rho_{\rm r} = \rho_{\rm r,0}(1+z)^4$, and $\Omega_{\rm i,0} = \frac{\rho_{\rm i,0}}{\rho_{\rm c,0}}$, we finally obtain
\begin{equation}
\left(\frac{H}{H_0}\right)^2 = \alpha + (\Omega_{\rm dm,0} - \alpha)(1+z)^3 + \Omega_{\rm b,0}(1+z)^3 + \Omega_{\rm r,0}(1+z)^4.
\label{A7}
\end{equation}

\section{Model II}\label{appen-B}

The continuity equation for the dark matter energy density $\rho_{\rm dm}$ in Model II is given by
\begin{equation}
\dot{\rho}_{\rm dm} + 3H\rho_{\rm dm} = 3\alpha H \left(\frac{\rho_{\rm c,0}}{\rho_{\rm dm}}\right)^l \rho_{\rm dm}.
\label{B1}
\end{equation}

Using the variable $N = \ln a$ and the relation $H = \frac{1}{a}\frac{da}{dt}$, Eq.~(\ref{B1}) reduces to
\begin{equation}
\frac{d\rho_{\rm dm}}{dN} + 3\rho_{\rm dm} = 3\alpha \rho_{\rm c,0}^l \rho_{\rm dm}^{\,1-l},
\label{B2}
\end{equation}
subject to the initial condition $\rho_{\rm dm}(N=0) = \rho_{\rm dm,0}$. This is a Bernoulli-type differential equation.

Multiplying Eq.~(\ref{B2}) by $\rho_{\rm dm}^{\,l-1}$, we obtain
\begin{equation}
\rho_{\rm dm}^{\,l-1}\frac{d\rho_{\rm dm}}{dN} + 3\rho_{\rm dm}^{\,l} = 3\alpha \rho_{\rm c,0}^l.
\end{equation}
Defining $y = \rho_{\rm dm}^{\,l}$, the above equation becomes
\begin{equation}
\frac{1}{l}\frac{dy}{dN} + 3y = 3\alpha \rho_{\rm c,0}^l,
\end{equation}
or equivalently,
\begin{equation}
\frac{dy}{dN} + 3l\,y = 3\alpha l\,\rho_{\rm c,0}^l.
\end{equation}

The integrating factor is $e^{3lN}$. Multiplying through by this factor, we obtain
\begin{equation}
\frac{d}{dN}\left(y e^{3lN}\right) = 3\alpha l\,\rho_{\rm c,0}^l e^{3lN}.
\end{equation}
Integrating, we find
\begin{equation}
y e^{3lN} = \frac{3\alpha l\,\rho_{\rm c,0}^l}{3l} e^{3lN} + C,
\end{equation}
which simplifies to
\begin{equation}
y = \alpha \rho_{\rm c,0}^l + C e^{-3lN}.
\end{equation}

Using the initial condition $y_0 = \rho_{\rm dm,0}^l$ at $N=0$, we obtain
\begin{equation}
C = \rho_{\rm dm,0}^l - \alpha \rho_{\rm c,0}^l.
\end{equation}
Thus,
\begin{equation}
\rho_{\rm dm}^l = \alpha \rho_{\rm c,0}^l + (\rho_{\rm dm,0}^l - \alpha \rho_{\rm c,0}^l)e^{-3lN}.
\end{equation}
Since $e^{-N} = a^{-1} = 1+z$, we finally obtain
\begin{equation}
\rho_{\rm dm} = \left[\alpha \rho_{\rm c,0}^l + (\rho_{\rm dm,0}^l - \alpha \rho_{\rm c,0}^l)(1+z)^{3l}\right]^{1/l}.
\end{equation}

Dividing by $\rho_{\rm c,0}$ and using $\Omega_{\rm dm,0} = \frac{\rho_{\rm dm,0}}{\rho_{\rm c,0}}$, we get
\begin{equation}
\frac{\rho_{\rm dm}}{\rho_{\rm c,0}} = \left[\alpha + (\Omega_{\rm dm,0}^l - \alpha)(1+z)^{3l}\right]^{1/l}.
\label{B3}
\end{equation}

Using Eq.~(\ref{A6}) along with $\rho_{\rm b} = \rho_{\rm b,0}(1+z)^3$ and $\rho_{\rm r} = \rho_{\rm r,0}(1+z)^4$, we obtain
\begin{align}
\left(\frac{H}{H_0}\right)^2 =&
\left[\alpha + (\Omega_{\rm dm,0}^l - \alpha)(1+z)^{3l}\right]^{1/l}
+ \Omega_{\rm b,0}(1+z)^3\nonumber\\&+ \Omega_{\rm r,0}(1+z)^4.
\end{align}


\bibliography{biblio}

\end{document}